\documentstyle[epsf]{article}
\newcommand{\tresj}[6]{ \left( \begin{array}{ccc}
                              \textstyle #1 & #2 & #3 \\
                              \textstyle #4 & #5 & #6 
                             \end{array}
                        \right) } 
\newcommand{\seisj}[6]{ \left\{ \begin{array}{ccc}
                                #1 & #2 & #3 \\
                                #4 & #5 & #6 
                               \end{array}
                        \right\} } 
\newcommand{\nuevej}[9]{ \left\{ \begin{array}{ccc}
                                 #1 & #2 & #3 \\
                                 #4 & #5 & #6 \\
                                 #7 & #8 & #9 
                                \end{array}
                         \right\} } 
\newcounter{pepe}
{\end{eqnarray}%
\setcounter{equation}{\arabic{pepe}}%
}
\newcommand{\Db}{{\cal D}}
\newcommand{\Jb}{{\cal J}}
\newcommand{\Mb}{{\cal M}}
\newcommand{\Rb}{{\cal R}}

\newcommand{\nk}{{\bf k}}
\newcommand{\np}{{\bf p}}
\newcommand{\nq}{{\bf q}}
\newcommand{\nr}{{\bf r}}

\newcommand{\nw}{{\bf w}}

\newcommand{\nM}{{\bf M}}
\newcommand{\nX}{{\bf X}}

\newcommand{\nP}{{\bf P}}
\newcommand{\hp}{{\bf \hat{p}}}

\newcommand{\nsigma}{\mbox{\boldmath$\sigma$}}
\newcommand{\nOmega}{\mbox{\boldmath$\Omega$}}
\begin{document}
\begin{titlepage}
\mbox{}
\vspace*{2.5\fill}

{\Large\bf 
\begin{center}
%**************************************
%
  Spin Observables in Coincidence \\
  Electron Scattering from Nuclei I:\\
  Reduced Response Functions$^*$
%
%**************************************
\end{center}
}

\vspace{1\fill}

\begin{center}
%**************************************
{\large  J.E. Amaro$^{1,2}$ and 
         T.W. Donnelly$^1$} 
%**************************************
\end{center}

{\small
\begin{center}
$^1$ 
{\em Center for Theoretical Physics, Laboratory for Nuclear
     Science and Dept. of Physics,
}\\
{\em  Massachusetts Institute of Technology, Cambridge, MA 02139, U.S.A.
}\\[2mm]
$^2$ 
{\em Departamento de F\'{\i}sica Moderna, Universidad de Granada,
      Granada 18071, Spain
}
\end{center}
}

\kern 1.5 cm

\hrule
\kern 3mm

{\small
\noindent
%**************************************
{\bf Abstract}
%**************************************
\vspace{3mm}

A theoretical description of nucleon knockout reactions initiated by
polarized electron scattering from polarized nuclei is presented. Explicit 
expressions for the complete set of reduced response functions (independent
of the polarization angle) that can be experimentally obtained assuming
plane
waves for the electron are given in a general multipole expansion. The 
formalism is applied to the particular case of closed-shell-minus-one
nuclei 
using two models for the ejected nucleon, including the final-state
interaction 
phenomenologically with a complex optical potential and in the factorized 
plane-wave impulse approximation. Relativistic effects in the kinematics
and 
in the electromagnetic current are incorporated throughout --- specifically
a 
new expansion of the electromagnetic current in powers only of the struck 
nucleon momentum is employed. Results are presented for the nucleus
$^{39}$K. 
}

\kern 3mm
\hrule

\vfill
{\small 
$^*$ This work is supported in part by funds provided by the U.S. 
Department of Energy (D.O.E.) under cooperative agreement 
\#DE-FC01-94ER40818, in part by D.G.E.S. (Spain) under 
grant No. PB95-1204 and the Junta de Andaluc\'{\i}a (Spain)
and in part by NATO Collaborative Research Grant \#940183.
}

\noindent MIT/CTP\#2600 \hfill March 1997

\end{titlepage}

%**************************************

\section{Introduction}

%**************************************

The completely unpolarized quasi-free $(e,e'p)$ reaction
has been systematically used to probe single-particle properties 
in complex nuclei, such as momentum distributions and spectroscopic 
factors \cite{Fru84}, and the reliability of the spectroscopic information
deduced from these data is based on the weak dependence of
the extracted momentum distributions upon the electron scattering
kinematical 
conditions of the experiment \cite{Kel96}. This is presumed to be
especially
true in the quasielastic region, namely at sufficiently high momentum
transfer
$q$ and energy transfer $\omega\sim \sqrt{q^2+M^2}-M$ (with $M$ the nucleon
mass) where the process is dominated by the interaction between the
electron and a single nucleon in the target, final-state interactions (FSI)
are assumed to play as small a role as they can and the plane-wave impulse 
approximation (PWIA) should be expected to become at least roughly valid. 
Thus, to the extent that this
 simple picture applies, the knockout cross section can be factorized 
into the product of the electron-nucleus cross section and the
spectral function, although in analyzing experiments one must usually take 
into account at least some aspects of the FSI.

The last is certainly the situation  when polarized electrons 
are used (but without hadronic polarizations --- see later), since 
the beam analyzing power involves a new observable --- the fifth 
response function --- that vanishes in PWIA and accordingly depends 
fundamentally upon the presence of FSI \cite{Don84,Ras89}. Something 
similar happens when polarized nuclei are used as targets. In that 
case a total of nine classes of structure functions can be obtained from 
$(\vec{e},e'N)$ reactions, specific ones of these being zero in PWIA 
\cite{Cab93}. This means that some of the polarization observables are
expected to be sensitive to specific aspects of the reaction mechanism 
that are less pronounced in the unpolarized cross section. 
It is then of interest to study which are the relevant observables 
in these kinds of reactions and  to explore their importance and
sensitivity to different aspects of the modeling in the kinematical ranges 
of interest.  

In this paper we focus our attention on exclusive quasielastic
scattering of polarized electrons from polarized nuclei. 
First, based on the general formalism for (polarized)
coincidence reactions presented in Ref.~\cite{Ras89}
we specialize to the reaction $\vec{A}(\vec{e},e'N)B$
and, as an extension of our recent work on the inclusive
reaction \cite{Ama97}, we introduce in Sec.~2 a set of reduced
response functions which characterize the polarized cross section 
and which do not depend on the polarization angles,
but only on the momentum and energy transfer
$(q,\omega)$ and the missing momentum $p$ of the nucleon, having fixed the 
missing energy to knockout of a nucleon from a specific shell.
To do this, we perform a multipole expansion of the nuclear 
states and electromagnetic current. This involves a sum over third
components
of the angular momentum that is performed analytically.
The 
resulting exclusive cross section and hadronic structure functions 
also depend on the polarization and emission angles, which we can write 
explicitly as a linear combination of spherical harmonics. 
In particular, a multipole expansion in terms of
spherical harmonics of the polarization angles is obtained. 
The coefficients in this expansion are directly
related to what we call angular reduced response functions. These functions
constitute the basic ingredients which enter in the total cross section. 
We illustrate in Appendix A how to obtain the final multipole
expansion for the particular case of the longitudinal response.

Although we shall see that the number of angular reduced
response functions is very large, too much so to hope for 
experimental extraction of all of them in the general case, it is
of theoretical interest to study their sensitivity to specific
details of the reaction mechanism with the goal of deciding which are the
most relevant to measure. This is the subject of the present paper.

We explore these ideas within the context of the shell model. In Sec.~3
we specialize the general expressions to the particular 
case of nucleon knockout from closed-shell-minus-one nuclei,
leaving the daughter nucleus in a discrete state which  is described by 
two holes in the core. We apply the formalism to the $^{39}$K nucleus
which is described as a $d_{3/2}$ hole in the $^{40}$Ca closed shell core.
Our theoretical framework addresses the
final-state interaction of the ejected nucleon with the residual
nucleus by solving the Schr\"odinger equation with a phenomenological
(complex) optical potential. The main result of this section is that 
in the extreme shell model the reduced nuclear responses are 
proportional to the polarized response functions for a single 
polarized particle in a shell. This is more or less obvious for 
one-particle nuclei, but not for the case of one-hole nuclei, because 
the involved shell and target nucleus can have different angular momenta, and
the final daughter nucleus remains in a two-hole state with definite total 
angular momentum and then both holes are partially polarized. 
We prove the above result analytically for hole-nuclei by performing 
the sum over the total angular momenta of the final hadronic state
composed of the daughter nucleus plus the ejected particle (see Appendix
B).

One of the aims of our systematic investigation of spin 
observables initiated in this paper is to determine whether momentum
distributions of polarized nuclei (in particular nuclei near closed shells)
may look different from those in the unpolarized case.
In fact, single-nucleon knockout from polarized nuclei can 
provide a new probe of the spin-dependent nuclear spectral 
function which represents the probability of finding
a nucleon in the target with given energy, momentum and spin projection.
Then, spin observables can be used to explore the spin distribution
of the single-particle orbitals, from which one can  extract 
valuable information about the complete spatial distribution 
of the orbits. The general framework for such studies in PWIA
was provided in Refs.~\cite{Cab93},\cite{Cab94},
and it was applied in Ref.~\cite{Ama97} to the study
of the inclusive polarized responses of one-hole nuclei.
Accordingly in Sec.~4 we introduce the formalism 
needed for the particular case of one-hole nuclei. 

The factorized PWIA gives us a very clear picture of the initial state 
physics, but the 
final-state propagation of the outgoing nucleon has to be taken into
account. 
In the present work we are able to go beyond PWIA and to determine
to what degree the FSI effects can obscure the extraction of those 
single-particle properties. An exploratory study to set the scale of
possible 
measurable quantities including the FSI was performed in Ref.~\cite{Bof88}.

It is important to note that the quasi-free conditions that favour the 
validity 
of the PWIA require a high value of the momentum transfer $q$ which
prohibits 
the usual non-relativistic expansions of the electromagnetic current in
powers
of $q/M$. Furthermore, for these kinematics the ejected nucleon
becomes relativistic. Thus, in this work we use a new approximation
to the on-shell relativistic one-body current that was tested 
in Ref.~\cite{Ama96}; this involves an expansion only in 
powers of $\eta=p/M$, but not in $\kappa=q/2M$ or $\lambda=\omega/2M$.
In addition we incorporate relativistic kinematics throughout in the 
calculations.

%*************************************************************

\section{General formalism for exclusive electron scattering}

%*************************************************************

%------------------------------------------------------------
\subsection{Cross section}
%------------------------------------------------------------

We begin this section by considering the scattering of polarized
electrons from polarized targets.
The general formalism for coincidence electron scattering in which  all of
the particles involved can be polarized has been presented in
Ref.~\cite{Ras89}. In the present treatment we have re-worked this formalism 
keeping in mind the specific process in which only the initial particles 
are polarized and written down the general equations in a 
way that is  oriented towards studies of nucleon knockout reactions.
For example, for the shell model applications considered here, it is 
convenient to use $jj$ coupling instead of $LS$ coupling as was used in 
Ref.~\cite{Ras89}. Also, following the scheme introduced in
Ref.~\cite{Don86}, 
we make an expansion of the nuclear structure functions 
in a basis of spherical harmonics of the polarization angles,
in that way introducing a set of reduced response functions that completely
determine the process.

Insofar as the electron scattering process is concerned, as discussed in 
Ref.~\cite{Ras89} we limit our 
attention to the one-photon-exchange or plane-wave Born approximation
(PWBA). 
The four momenta of the incident and scattered
electrons 
are labelled $K^{\mu}=(\epsilon_e,\nk_e)$ and 
$K'{}^{\mu}=(\epsilon'_e,\nk'_e)$, respectively. The four-momentum 
transfer is given by $Q^{\mu}= ( \omega,\nq)$. We work in the laboratory
system, where the initial nucleus is at rest in the state denoted
by $|A\rangle$. After the interaction a proton or neutron with momentum 
direction $\hp'$ and energy $E'$ is detected and the (undetected) daughter 
nucleus is left in a discrete state $|B\rangle$. We can assume that when
the 
nucleon is detected in coincidence with the scattered electron, it is 
essentially at infinity, and therefore on-shell, so that the magnitude of
the 
asymptotic value of the momentum $p'$ is related to the energy by 
$E'=\sqrt{{p'}^2+M^2}$, as usual for free particles, although at short 
distances when the effects of the FSI are important, the momentum of the 
ejected nucleon is not well-defined. Therefore, we define the momentum
$\np'$ of the ejectile as {\em the well-defined asymptotic value} and so 
can label the nucleon wave function as 
$|\np'\rangle \equiv |E',\hp'\rangle$. 

On the other hand,  later in using  the formalism in the shell model where 
recoil is not handled properly we make no attempt to extract the center of
mass 
motion from the nuclear wave function. Note that this is not a real
limitation 
of the general formalism that follows, although a proper treatment of the 
problem is far from trivial. Indeed, an artificial extraction {\em by hand}
of 
the center of mass motion in the shell model is known  to generate spurious
effects that only can be controlled in very specific cases, in particular
not 
in the conventional continuum shell model. Our expectations here in
applying
the shell model are that the recoil effects in the quasi-free regime are
not 
especially important for nuclei as heavy as $^{39}$K. 

Also, for the moment only the electron is treated relativistically. Later 
we will approximately include the relativistic effects in the ejectile
nucleon
by using relativistic kinematics and an appropriate electromagnetic current
operator. Therefore, the $S$-matrix element for the above process
can be written as
\begin{equation}
S_{fi} = -2\pi i \delta(E'+E_B-E_A-\omega)
         \frac{4\pi\alpha}{Q^2}
         \langle K'h'|j_{\mu}(0)|Kh\rangle
         \langle \np'm_s,B|J^{\mu}(\nq)|A\rangle,
\end{equation}
where $h$ and $h'$ are the helicities of the (ultra-relativistic) initial
and 
final electrons, $E_A$ and $E_B$ are the energies of the initial and
daughter 
nuclei, $Q^2=\omega^2-q^2\leq 0$ is the spacelike four-momentum transfer, 
$\alpha$ is the fine structure constant, $j_{\mu}(0)$ is the 
electromagnetic 
current operator for the electron, $m_s$ is the spin projection of the 
ejectile, and $J^{\mu}(\nq)$ is the Fourier transform of the nuclear 
electromagnetic current operator. The differential cross section in the 
laboratory system is proportional to 
$|S_{fi}|^2$ and to the electron-nucleon phase space
$\frac{Vd^3k'_e}{(2\pi)^3}  {\rm d}E'\,{\rm d}\hp'$, 
where ${\rm d}\hp'={\rm d}\phi'{\rm d}\cos\theta'$ 
is the solid angle differential corresponding to the emission angles
$(\theta',\phi')$. 
Performing the  sum over final undetected electron helicity 
and  integrating over the  nucleon energy 
$E'$ using the energy conservation condition $E'=E_A+\omega-E_B$, 
the resulting cross section
is proportional to the contraction 
$\eta_{\mu\nu}W^{\mu\nu}$
of the leptonic and hadronic tensors
$\eta_{\mu\nu}$ and  $W^{\mu\nu}$ 
given in Ref.~\cite{Don86} and below, respectively.
In practice, the ejectile energy $E'$ is detected, so the integration
over the energy corresponds to experimentally suming up all of the events
that contribute to the selected peak in the missing energy spectrum. 

Finally, the contraction $\eta_{\mu\nu}W^{\mu\nu}$ is performed as usual 
in the extreme relativistic limit for the electron and in the coordinate 
system with the $z$-axis along $\nq$ and the $x$-axis in the scattering
plane. Using current conservation, the result can be written
\cite{Don86}
\begin{eqnarray}
\frac{d\sigma}{d\epsilon'_ed\Omega'_e d\hp'}&=&
\sigma_M\left[v_L\Rb^L + v_T\Rb^T + v_{TL}\Rb^{TL}+v_{TT}\Rb^{TT}
              +h(v_{T'}\Rb^{T'}+v_{TL'}\Rb^{TL'})
        \right] \\
&=& \Sigma+h\Delta, \label{sigmaa}
\end{eqnarray}
where 
$\sigma_M$
 is the Mott cross section,
The electron kinematic
 factors $v_K$ are defined in Ref. \cite{Don86}.
The six exclusive nuclear response functions  $\Rb^{K}$ 
are the components given in Eqs.~(\ref{22}--\ref{27}) below
of the hadronic tensor for the exclusive process defined by
\begin{equation}
W^{\mu\nu}= \sum_{m_sM_B}
           \langle \np'm_sB|\hat{J}^{\mu}({\bf q})|A\rangle^*
           \langle \np'm_sB|\hat{J}^{\nu}({\bf q})|A\rangle,
\end{equation}
where we sum over the undetected polarizations of the final 
particles. For a treatment of the general case where all of the 
particles are polarized, see Ref.~\cite{Ras89}. Here we have followed 
a somewhat different derivation that is more appropriate for the particular
application to the shell model. 

%-------------------------------------
\subsection{Polarized response functions}
%-------------------------------------

%------------------------------------------------------------
\subsubsection{Hadronic final states}
%------------------------------------------------------------

Now some more detailed specifications of the nuclear states themselves are
in 
order. The initial nuclear state is assumed to be 100\% polarized relative
to 
an arbitrary axis of quantization whose orientation is specified by the
angles
$\Omega^*=(\theta^*,\phi^*)$ relative to the coordinate system defined
above. 
We also denote by $\nOmega^*$ the unit vector pointing in the $\Omega^*$ 
direction. Therefore, $\theta^*$ is the angle between $\nq$ and
$\nOmega^*$; note that the conventions here have $\phi^*$ 
with respect to the electron plane, whereas Ref.~\cite{Ras89} starts 
with reference to the hadron plane.
We express the nuclear wave function in terms of state vectors defined with
respect to the $z$-axis:
\begin{equation} \label{17}
|A\rangle= |J_iJ_i(\Omega^*)\rangle
         = R(\phi^*,\theta^*,0)|J_iJ_i\rangle
         = \sum_{M_i} \Db^{(J_i)}_{M_iJ_i}(\Omega^*)|J_iM_i\rangle.
\end{equation}  
Note also that the three Euler angles $(\phi^*,\theta^*,0)$ of the rotation
matrix 
are different from those specified in Ref.~\cite{Ras89}. However, since any
rotation around the $\Omega^*$ axis does not modify the 
polarization vector, the resulting observables are exactly the
same in the two approaches.

The final hadronic state corresponds asymptotically to a 
nucleon with energy $E'$, momentum direction $\hp'$ with angles
$(\theta',\phi')$ referred to the same coordinate system, 
and with (unobserved) spin projection $m_s$, together with an 
(undetected) residual  nucleus in the discrete state $|B\rangle$
(that in the general case is recoiling with momentum $\nP_B$).
In analogy with the plane-wave expansion in spherical harmonics,
we assume that a similar expansion can be performed for the hadronic
states
\begin{equation} \label{19}
|\np'm_sB\rangle= \sum_{lMjmJ_fM_f}i^lY^*_{lM}(\hp')
\textstyle                \langle \frac12 m_s lM|jm\rangle
                  \langle jmJ_BM_B |J_fM_f\rangle
                  |(E'lj)J_B;J_fM_f \rangle.
\end{equation}
Here $ |(E'lj)J_B;J_fM_f \rangle$ is a continuum nuclear state having 
asymptotic behaviour
\begin{equation} \label{20}
  |(E'lj)J_B;J_fM_f \rangle \longrightarrow 
  \left[ a^{\dagger}_{E'lj}|J_B\rangle
  \right]_{J_fM_f},
\end{equation}
where $ a^{\dagger}_{E'lj}$ creates a single nucleon with energy 
$E'$ and angular momenta $(lj)$. Note that the condition in Eq.~(\ref{20}) 
only has to be verified asymptotically. Therefore we do not make any 
assumption about the particular nuclear model to be used until later,
keeping
the following steps in this section completely general. However, it may be 
helpful to anticipate the modeling used later by noting that any particular
theory of the reaction mechanism has to be able to provide a suitable form
for the continuum states $  |(E'lj)J_B;J_fM_f \rangle$. In particular, 
in the simple case of the extreme shell model, Eq.~(\ref{20}) 
is assumed to be true not only asymptotically, but also at short 
distances. In the simplest case of the PWIA the radial wave function
is proportional to a spherical Bessel function and then Eq.~(\ref{19})
reduces  to a plane wave times the daughter state $|B\rangle$.
As for the normalization of the states, it is
\begin{equation}\label{21}
  \langle(E'l'j')J'_B;J'_fM'_f |(Elj)J_B;J_fM_f \rangle 
= \delta(E-E')\delta_{ll'}\delta_{jj'}\delta_{J_BJ'_B}
  \delta_{J_fJ'_f}\delta_{M_fM'_f}.
\end{equation}
This relation, together with the asymptotic condition in Eq.~(\ref{20}) 
uniquely determines the nuclear states. 

%-----------------------------------------
\subsubsection{Multipole analysis}
%-----------------------------------------

In order to perform a multipole expansion of the responses, it is
convenient to write them in terms of the spherical 
tensor components of the current \cite{Ras89}:
\begin{eqnarray}
\Rb^L    &=& W^{00}=\sum\rho^*\rho \label{22}\\
\Rb^T    &=& W^{xx}+W^{yy}=\sum[|J_{-1}|^2+|J_{+1}|^2] \label{13}\\
\Rb^{TT} &=& W^{yy}-W^{xx}
          =  \sum[J_{-1}^*J_{+1} + J_{+1}^*J_{-1}] \label{componentes2}\\
\Rb^{TL} &=& \sqrt{2}(W^{0x}+W^{x0})=
             -\sum 2\mbox{Re}[\rho^*(J_{+1}-J_{-1})]\\
\Rb^{TL'}&=& i\sqrt{2}(W^{0y}-W^{y0})
             = -\sum 2\mbox{Re}[\rho^*(J_{+1}+J_{-1})] \label{15}\\
\Rb^{T'} &=& i(W^{xy}-W^{yx})
          =  \sum[|J_{+1}|^2-|J_{-1}|^2]. \label{27}
\end{eqnarray}
Now we perform the usual multipole expansion for the 
charge and transverse current operators:
\begin{eqnarray}
\hat{\rho}({\bf q}) &=& \sqrt{4\pi}\sum_J i^J[J] 
                        \hat{M}_{J0}(q)\label{28}\\
\hat{J}_{m}({\bf q})&=& -\sqrt{2\pi}\sum_J i^J[J]
                \left[ \hat{T}^{el}_{Jm}(q)+m\hat{T}^{mag}_{Jm}(q)\right],
\end{eqnarray}
where $m$ is the index of the spherical components and $\hat{M}_{Jm}$,
$\hat{T}^{el}_{Jm}$ and $\hat{T}^{mag}_{Jm}$ are the usual Coulomb (CJ), 
electric (EJ) and magnetic (MJ) multipoles. Throughout we use the notation 
$[J]\equiv\sqrt{2J+1}$ for any angular momentum variable $J$.

Using the above expansion, the nuclear tensor can be written as a sum
of factors of the type
\begin{equation} \label{30}
B^{m'm}_{J'J} =  \sum_{m_sM_B}
                 \langle \np'm_s,J_BM_B|\hat{T}'_{J'm'}|A\rangle^*
                 \langle \np'm_s,J_BM_B|\hat{T}_{Jm}|A\rangle,
\end{equation}
where $\hat{T}$ and $\hat{T}'$ are any of the $C$, $E$ or $M$ operators.
For instance, in Appendix A we show the expression
for the $L$ response given in terms of the $B^{00}_{J'J}$ coefficients
(see Eq.~(\ref{app188})).
The parameters $m$ or $m'$ take on the value 0 when the
electromagnetic operator involved in the corresponding response
is the charge one, while they take on the values 1 or -1 when
transverse current components occur.
Inserting the expressions Eqs.~(\ref{17},\ref{19}) of the
initial and final nuclear states, respectively, into Eq. (\ref{30}), 
we need to evaluate 
the following multiple sum
\begin{eqnarray}
B^{m'm}_{J'J} &=& \sum_{m_sM_B}
                \sum_{l'M'}
                \sum_{j'm'_p}
                \sum_{J'_fM'_f}
                \sum_{lMjm_p}
                \sum_{J_fM_f}
                \sum_{M_iM'_i}
                i^{l'-l}Y_{lM}(\hp')Y^*_{l'M'}(\hp')
                \Db^{(J_i)*}_{M'_iJ_i}(\Omega^*)
                \Db^{(J_i)}_{M_iJ_i}(\Omega^*) \nonumber\\
&&      \times  \langle {\textstyle\frac12}m_sl'M'|j'm'_p\rangle
                \langle j'm'_pJ_BM_B|J'_fM'_f\rangle
                \langle {\textstyle\frac12}m_slM|jm_p\rangle
                \langle jm_pJ_BM_B|J_fM_f\rangle \nonumber\\
&&      \times 
                \langle (l'j')J_B,J'_fM'_f|\hat{T}'_{J'm'}|J_iM'_i\rangle^*
                \langle (lj)J_B,J_fM_f|\hat{T}_{Jm}|J_iM_i\rangle.
        \label{31}
\end{eqnarray}
The procedure to simplify the above expression is then the following.

\begin{enumerate}

\item 
Write the product of the two spherical harmonics 
$Y_{lM}Y^*_{l'M'}$
as a linear combination of spherical harmonics
$  Y_{\Jb'\Mb'}(\hp')$. Therefore the index $\cal J'$ below refers to 
the multipole expansion in emission angles.

\item 
Rewrite the dependence on the polarization direction in
the product of two rotation matrices 
${\cal D}^{(J_i)*}_{M'_iJ_i}(\Omega^*){\cal D}^{(J_i)}_{M_iJ_i}(\Omega^*)$
by developing this in spherical
harmonics of the polarization angles $Y_{\cal JM}(\Omega^*)$,
and where the coefficients in the expansion  also contain as
a factor 
the Fano tensor for 100\% polarization 
given by $f^i_{\cal J}=\langle J_iJ_iJ_i-J_i|{\cal J}0\rangle$ (see, for 
example, Ref.~\cite{Ras89}). Therefore, the indices 
${\cal JM}$ in the equations and reduced response functions 
defined below refer to this multipole expansion of the polarization
dependence.

\item  Use the Wigner-Eckart theorem in the matrix elements.
\end{enumerate}

After these steps, all of the dependence upon third components 
is via 3-$j$ coefficients, and the corresponding sums can be 
performed analytically using Racah algebra, but
here we only quote the final result. 
Also, following Ref.~\cite{Ras89}, in order to simplify the notation 
it is convenient to define  multiple indices $\sigma$, $\sigma'$ by
$\sigma\equiv \{l,j,J_f,J\}$ and 
$\sigma'\equiv \{l',j',J'_f,J'\}$, 
and a coefficient
\begin{eqnarray}
\Phi_{\sigma'\sigma}(\Jb,\Jb',L;J_i\rightarrow J_B) &\equiv&
        [J][J'][j][j'][J_f][J'_f][\Jb'][L]
        (-1)^{J'+J_B+J_f+1/2+\Jb+\Jb'} \nonumber\\
&& \kern -1.5cm \times
        \tresj{j'}{j}{\cal J'}{\textstyle\frac12}{-\textstyle\frac12}{0}
        \seisj{j'}{j}{\cal J'}{J_f}{J'_f}{J_B}
        \nuevej{J}{J'}{L}{J_i}{J_i}{\cal J}{J_f}{J'_f}{\cal J'}.
        \label{35}
\end{eqnarray}
As  the values of $J_i$ and $J_B$ are fixed, we shall usually not write
them 
in the argument of $\Phi_{\sigma'\sigma}$. We quote here an important 
symmetry property of this coefficient under exchange of the two indices
$\sigma'\sigma$ that can be easily verified:
\begin{equation}\label{36}
\Phi_{\sigma'\sigma}(\Jb\Jb'L)
=(-1)^{\Jb+\Jb'+L}
\Phi_{\sigma\sigma'}(\Jb\Jb'L).
\end{equation}

With these definitions, the general expression for $B^{m'm}_{J'J}$ becomes
the following:
\begin{eqnarray}
[J][J']B^{m'm}_{J'J} &=& \sum_{ljJ_f} \sum_{l'j'J'_f}
                        \sum_{\cal JJ'}\sum_{LM}
                        i^{l'-l}P^+_{l+l'+\cal J'}
                        (-1)^m f^i_{\cal J}
                        [Y_{\cal J}(\Omega^*)Y_{\cal J'}(\hp')]_{LM}
                        \nonumber\\
        & & \times      \Phi_{\sigma'\sigma}(\Jb\Jb'L)
                        \tresj{J'}{J}{L}{m'}{-m}{M}
                        \langle f'\|\hat{T}'_{J'}\| J_i \rangle^*
                        \langle f\|\hat{T}_{J}\| J_i \rangle,
        \label{37}
\end{eqnarray}
where we define parity projectors 
$P^{\pm}_{J}= (1\pm (-1)^J)/2$
and use  shorthand for the generic final states,
$|f\rangle = |(E'lj)J_B;J_f\rangle$
and
$|f'\rangle = |(E'l'j')J_B;J'_f\rangle$.

Now, in order to write compact expressions for the different response 
functions it is convenient to define the following Coulomb,
electric and magnetic  multipole matrix 
elements,
\begin{eqnarray}
C_{\sigma} &\equiv& \langle f \| \hat{M}_J \| J_i\rangle\label{40}\\
E_{\sigma} &\equiv& \langle f \| \hat{T}^{el}_J \| J_i\rangle\\
M_{\sigma} &\equiv& \langle f \| i\hat{T}^{mag}_J \| J_i\rangle,
\end{eqnarray}
that in general are complex functions only of $(q,\omega)$, since the 
asymptotic value of the energy is given by $E'=E_A+\omega-E_B$, and
both $E_A$ and $E_B$ are fixed. The 
response functions are defined through linear
combinations of the real and imaginary parts of the
following quadratic products of the 
multipole matrix elements:
\begin{eqnarray}
R^L_{\sigma'\sigma}+iI^L_{\sigma'\sigma}
&=& 
C_{\sigma'}^*C_{\sigma} 
\label{43}\\
R^{T1}_{\sigma'\sigma}+iI^{T1}_{\sigma'\sigma}
&=& 
E_{\sigma'}^*E_{\sigma}+ 
M_{\sigma'}^*M_{\sigma} 
\label{44}\\
R^{T2}_{\sigma'\sigma}+iI^{T2}_{\sigma'\sigma}
&=& 
E_{\sigma'}^*M_{\sigma}- 
M_{\sigma'}^*E_{\sigma} 
\label{45}\\
R^{TL1}_{\sigma'\sigma}+iI^{TL1}_{\sigma'\sigma}
&=& 
C_{\sigma'}^*E_{\sigma} 
\label{46}\\
R^{TL2}_{\sigma'\sigma}+iI^{TL2}_{\sigma'\sigma}
&=& 
C_{\sigma'}^*M_{\sigma} 
\label{47}\\
R^{TT1}_{\sigma'\sigma}+iI^{TT1}_{\sigma'\sigma}
&=& 
E_{\sigma'}^*E_{\sigma}-
M_{\sigma'}^*M_{\sigma} 
\label{48}\\
R^{TT2}_{\sigma'\sigma}+iI^{TT2}_{\sigma'\sigma}
&=& 
E_{\sigma'}^*M_{\sigma}+ 
M_{\sigma'}^*E_{\sigma} ,
\end{eqnarray}
where the functions 
$R^K_{\sigma'\sigma}$ and
$I^K_{\sigma'\sigma}$ 
are real functions only of $(q,\omega)$.
Also it is convenient to write the angular dependence in Eq.~(\ref{37})
as a sum of real plus imaginary parts. Thus we define the real functions
$A_{\Jb\Jb'LM}(\Omega^*,\hp')$ and
$B_{\Jb\Jb'LM}(\Omega^*,\hp')$
as
\begin{equation}\label{50}
A_{\Jb\Jb'LM}(\Omega^*,\hp')
+iB_{\Jb\Jb'LM}(\Omega^*,\hp') = 
[Y_{\cal J}(\Omega^*)\otimes Y_{\cal J'}(\hp')]_{LM}.
\end{equation}
Finally we also define two  parity functions
\begin{equation}\label{51}
\xi_{J'J}^+\equiv (-1)^{(J'-J)/2}P^+_{J'+J}, \kern 1cm
\xi_{J'J}^-\equiv (-1)^{(J'-J+1)/2}P^-_{J'+J}.
\end{equation}

After these preliminary definitions and following a procedure 
similar to that in Ref.~\cite{Ras89} (summarized in Appendix A 
for the particular case of the $L$-response), we are then in a 
position to write the exclusive responses in the following 
relatively compact way:
\begin{eqnarray}
\Rb^L &=& 4\pi\sum_{\Jb\Jb'L}f^i_{\cal J}
        \left\{ 
                P^+_{\cal J} A_{\Jb\Jb'L0} W^{L(+)}_{\Jb\Jb'L}
                -P^-_{\cal J} B_{\Jb\Jb'L0} W^{L(-)}_{\Jb\Jb'L}
        \right\}
        \label{55}\\
\Rb^T &=& 4\pi\sum_{\Jb\Jb'L}f^i_{\cal J}P^+_{\Jb'+L}
        \left\{ 
                P^+_{\cal J} A_{\Jb\Jb'L0} W^{T(+)}_{\Jb\Jb'L}
                -P^-_{\cal J} B_{\Jb\Jb'L0} W^{T(-)}_{\Jb\Jb'L}
        \right\}
        \label{56}\\
\Rb^{T'} &=& 4\pi\sum_{\Jb\Jb'L}f^i_{\cal J}P^-_{\Jb'+L}
        \left\{ 
                P^-_{\cal J} A_{\Jb\Jb'L0} W^{T'(-)}_{\Jb\Jb'L}
                -P^+_{\cal J} B_{\Jb\Jb'L0} W^{T'(+)}_{\Jb\Jb'L}
        \right\}
        \label{57}\\
\Rb^{TL} &=& 4\pi\sum_{\Jb\Jb'L}f^i_{\cal J}
        \left\{ 
                P^+_{\cal J} A_{\Jb\Jb'L1} W^{TL(+)}_{\Jb\Jb'L}
                -P^-_{\cal J} B_{\Jb\Jb'L1} W^{TL(-)}_{\Jb\Jb'L}
        \right\}
        \label{58}\\
\Rb^{TL'} &=& 4\pi\sum_{\Jb\Jb'L}f^i_{\cal J}
        \left\{ 
                P^-_{\cal J} A_{\Jb\Jb'L1} W^{TL'(-)}_{\Jb\Jb'L}
                -P^+_{\cal J} B_{\Jb\Jb'L1} W^{TL'(+)}_{\Jb\Jb'L}
        \right\}
        \label{59}\\
\Rb^{TT} &=& 4\pi\sum_{\Jb\Jb'L}f^i_{\cal J}
        \left\{ 
                P^+_{\cal J} A_{\Jb\Jb'L2} W^{TT(+)}_{\Jb\Jb'L}
                -P^-_{\cal J} B_{\Jb\Jb'L2} W^{TT(-)}_{\Jb\Jb'L}
        \right\}.
        \label{60}
\end{eqnarray}
The {\em reduced response functions} 
$W^{K(\pm)}_{\Jb\Jb'L}(q,\omega)$ contain all 
of the information about the nuclear structure and reaction mechanism 
and they are the following real quadratic forms involving the basic
(complex) 
multipoles
 $C_{\sigma}$,
 $E_{\sigma}$ and
 $M_{\sigma}$:
\begin{eqnarray}
W^{L(+)}_{\Jb\Jb'L} &=& \sum_{\sigma'\sigma}P^+_{l+l'+\Jb'}
                        \Phi_{\sigma'\sigma}
                        \tresj{J}{J'}{L}{0}{0}{0}
                        \xi^+_{J'-l',J-l} 
                        R^L_{\sigma'\sigma}
                        \label{61}\\
W^{L(-)}_{\Jb\Jb'L} &=& \sum_{\sigma'\sigma}P^+_{l+l'+\Jb'}
                        \Phi_{\sigma'\sigma}
                        \tresj{J}{J'}{L}{0}{0}{0}
                        \xi^+_{J'-l',J-l} 
                        I^L_{\sigma'\sigma}
                        \label{62}\\
W^{T(+)}_{\Jb\Jb'L} &=& -\sum_{\sigma'\sigma}P^+_{l+l'+\Jb'}
                        \Phi_{\sigma'\sigma}
                        \tresj{J}{J'}{L}{1}{-1}{0}
                        (\xi^+_{J'-l',J-l} 
                        R^{T1}_{\sigma'\sigma}
                        +\xi^-_{J'-l',J-l} 
                        R^{T2}_{\sigma'\sigma})
                        \label{63}\\
W^{T(-)}_{\Jb\Jb'L} &=& -\sum_{\sigma'\sigma}P^+_{l+l'+\Jb'}
                        \Phi_{\sigma'\sigma}
                        \tresj{J}{J'}{L}{1}{-1}{0}
                        (\xi^+_{J'-l',J-l} 
                        I^{T1}_{\sigma'\sigma}
                        +\xi^-_{J'-l',J-l} 
                        I^{T2}_{\sigma'\sigma})
                        \label{64}\\
W^{T'(-)}_{\Jb\Jb'L} &=& W^{T(+)}_{\Jb\Jb'L}
                        \label{65}\\
W^{T'(+)}_{\Jb\Jb'L} &=& W^{T(-)}_{\Jb\Jb'L}
                        \label{66}\\
W^{TL(+)}_{\Jb\Jb'L} &=& -2\sqrt{2}(-1)^{\Jb'+L}
                        \sum_{\sigma'\sigma}P^+_{l+l'+\Jb'}
                        \Phi_{\sigma'\sigma}
                        \tresj{J}{J'}{L}{0}{1}{-1}
                        \nonumber\\
        &&\times
                        (\xi^+_{J'-l',J-l} 
                        R^{TL1}_{\sigma'\sigma}
                        -\xi^-_{J'-l',J-l} 
                        R^{TL2}_{\sigma'\sigma})
                        \label{67}\\
W^{TL(-)}_{\Jb\Jb'L} &=& -2\sqrt{2}(-1)^{\Jb'+L}
                        \sum_{\sigma'\sigma}P^+_{l+l'+\Jb'}
                        \Phi_{\sigma'\sigma}
                        \tresj{J}{J'}{L}{0}{1}{-1}
                        \nonumber\\
        &&\times
                        (\xi^+_{J'-l',J-l} 
                        I^{TL1}_{\sigma'\sigma}
                        -\xi^-_{J'-l',J-l} 
                        I^{TL2}_{\sigma'\sigma})
                        \label{68}\\
W^{TL'(-)}_{\Jb\Jb'L} &=& W^{TL(+)}_{\Jb\Jb'L}
                        \label{69}\\
W^{TL'(+)}_{\Jb\Jb'L} &=& W^{TL(-)}_{\Jb\Jb'L}
                        \label{70}\\
W^{TT(+)}_{\Jb\Jb'L} &=& -\sum_{\sigma'\sigma}P^+_{l+l'+\Jb'}
                         (-1)^{\Jb'+L}
                        \Phi_{\sigma'\sigma}
                        \tresj{J}{J'}{L}{1}{1}{-2}
                        \nonumber\\
        &&\times
                        (\xi^+_{J'-l',J-l} 
                        R^{TT1}_{\sigma'\sigma}
                        -\xi^-_{J'-l',J-l} 
                        R^{TT2}_{\sigma'\sigma})
                        \\
W^{TT(-)}_{\Jb\Jb'L} &=& -\sum_{\sigma'\sigma}P^+_{l+l'+\Jb'}
                         (-1)^{\Jb'+L}
                        \Phi_{\sigma'\sigma}
                        \tresj{J}{J'}{L}{1}{1}{-2}
                        \nonumber\\
        &&\times
                        (\xi^+_{J'-l',J-l} 
                        I^{TT1}_{\sigma'\sigma}
                        -\xi^-_{J'-l',J-l} 
                        I^{TT2}_{\sigma'\sigma}).
                        \label{72}
\end{eqnarray}

%-------------------------------------------------------------
\subsection{Angular reduced response functions}
%-------------------------------------------------------------

The $(q,\omega)$-dependent reduced response functions discussed above
completely determine the responses; however, it is not always 
very convenient to work with them because their number is infinite.
In fact, the multipole expansion in Eq.~(\ref{19}) of the nuclear
wave function involves an infinite number of partial waves $l$.
There are interferences between different partial waves $l$ and $l'$
in the response functions, where $l$ and $l'$ are effectively 
coupled to total angular momentum $\Jb'$. As a consequence, 
the sum over $\Jb'$ in Eqs.~(\ref{55}-\ref{60}) 
is also infinite. The same happens for the photon angular momenta, 
labeled by the multipole values $J$, $J'$ coupled to 
$L$, that are also summed up to infinity. In practice, of course,  
we actually sum a (usually) large but finite number of terms until 
convergence is reached.   

For experimental separation purposes and also for theoretical analysis it is 
convenient to define an alternative finite set of 
what we call {\em angular reduced response
functions}  using the fact that the values of $\cal J$ run from 
$\Jb=0,\ldots,2J_i$.
Then  an expansion in Legendre functions $P_{\cal J}^{\cal M}(\cos\theta^*)$ 
and trigonometric functions of the angles $\phi^*$ and $\phi'$ has a finite
well-defined number of terms, from which a useful 
set of response functions (coefficients of the expansion) 
that  depend on $q$, $\omega$ and the emission 
angle $\theta'$ can be defined as some combinations of the functions
$W^{K(\pm)}_{\Jb\Jb'L}(q,\omega)$ and Legendre functions of the angle 
$\theta'$. To do this we note that all of the angular dependence of the 
response functions is through the coupling of two spherical harmonics that 
can be written as
\begin{equation}
[Y_{\cal J}(\Omega^*)Y_{\cal J'}(\hp')]_{LM}
= \sum_{\cal MM'}\langle{\cal JMJ'M'}|LM\rangle
  Y_{\cal JM}(\theta^*,0)Y_{\cal J'M'}(\theta',0)
  {\rm e}^{iM\phi'-i\Mb\Delta\phi},
\end{equation}
where $\Delta\phi\equiv \phi'-\phi^*$ is the azimuthal emission angle measured
with 
respect to the polarization projection on the scattering plane. Note that
the 
spherical harmonic for azimuthal angle equal to zero, 
$Y_{\cal JM}(\theta^*,0)$, is proportional to a Legendre function 
$P_{\cal J}^{\cal |M|}(\cos\theta^*)$. If we take the real and imaginary
parts 
of the above functions, as defined in Eq.~(\ref{50}), we have
\begin{eqnarray}
A_{\Jb\Jb'LM}(\Omega^*,\hp') 
        &=& \cos M\phi' \sum_{\Mb=0}^{\cal J}
                h_{\Jb\Jb'LM}^{\cal M}(\theta')
                P_{\cal J}^{\cal M}(\cos\theta^*)
                \cos(\Mb\Delta\phi) 
                \nonumber\\
        &&+ \sin M\phi' \sum_{\Mb=0}^{\cal J}
                \widetilde{h}_{\Jb\Jb'LM}^{\cal M}(\theta')
                P_{\cal J}^{\cal M}(\cos\theta^*)
                \sin(\Mb\Delta\phi)
                \label{76}\\ 
B_{\Jb\Jb'LM}(\Omega^*,\hp') 
        &=& \sin M\phi' \sum_{\Mb=0}^{\cal J}
                h_{\Jb\Jb'LM}^{\cal M}(\theta')
                P_{\cal J}^{\cal M}(\cos\theta^*)
                \cos(\Mb\Delta\phi) 
                \nonumber\\
        &&- \cos M\phi' \sum_{\Mb=0}^{\cal J}
                \widetilde{h}_{\Jb\Jb'LM}^{\cal M}(\theta')
                P_{\cal J}^{\cal M}(\cos\theta^*)
                \sin(\Mb\Delta\phi). 
                \label{77}
\end{eqnarray}
These equations can also be taken as the definitions of the angular
functions $h_{\Jb\Jb'LM}^{\cal M}(\theta')$ and
$\widetilde{h}_{\Jb\Jb'LM}^{\cal M}(\theta')$.
Inserting Eqs.~(\ref{76},\ref{77})
into Eqs.~(\ref{55}-\ref{60}) it is straightforward to write the 
response functions in the following way:
\begin{eqnarray}
\Rb^L &=& W^L \\
\Rb^T &=& W^T \\
\Rb^{TL} &=& \cos\phi' W^{TL}+\sin\phi' \widetilde{W}^{TL} \label{RTL}\\
\Rb^{TT} &=& \cos2\phi' W^{TT}+\sin2\phi' \widetilde{W}^{TT} \\
\Rb^{T'} &=& \widetilde{W}^{T'} \\
\Rb^{TL'} &=& \sin\phi' W^{TL'}+\cos\phi' \widetilde{W}^{TL'}.
\end{eqnarray}
The nine response functions $W^K$ and $\widetilde{W}^K$ without 
any sub-index depend on $q$, $\omega$, $\theta'$,
$\theta^*$, and $\Delta\phi$. 
Their explicit dependences on the angles $(\theta^*,\Delta\phi)$ are the
following:
\begin{eqnarray}
W^L &=& 4\pi\sum_{\Jb=0}^{2J_i}\sum_{\Mb=0}^{\cal J}
        f^i_{\cal J}P_{\cal J}^{\cal M}(\cos\theta^*)
        \left[
                P^+_{\cal J}c_{\cal M} W^{L(+)}_{\cal JM}
                +P^-_{\cal J}s_{\cal M}
                \widetilde{W}^{L(-)}_{\cal JM}
        \right]
        \label{64b}\\
W^T &=& 4\pi\sum_{\Jb=0}^{2J_i}\sum_{\Mb=0}^{\cal J}
        f^i_{\cal J}P_{\cal J}^{\cal M}(\cos\theta^*)
        \left[
                P^+_{\cal J}c_{\cal M} W^{T(+)}_{\cal JM}
                +P^-_{\cal J}s_{\cal M}
                \widetilde{W}^{T(-)}_{\cal JM}
        \right]
        \label{65b}\\
W^{TL} &=& 4\pi\sum_{\cal JM}
        f^i_{\cal J}P_{\cal J}^{\cal M}(\cos\theta^*)
        \left[
                P^+_{\cal J}c_{\cal M} W^{TL(+)}_{\cal JM}
                +P^-_{\cal J}s_{\cal M}
                \widetilde{W}^{TL(-)}_{\cal JM}
        \right]
        \label{66b}\\
\widetilde{W}^{TL} &=&
         4\pi\sum_{\cal JM}
        f^i_{\cal J}P_{\cal J}^{\cal M}(\cos\theta^*)
        \left[
                P^+_{\cal J}s_{\cal M}
                \widetilde{W}^{TL(+)}_{\cal JM}
                -P^-_{\cal J}c_{\cal M}
                W^{TL(-)}_{\cal JM}
        \right]
        \label{67b}\\
W^{TT} &=& 4\pi\sum_{\cal JM}
        f^i_{\cal J}P_{\cal J}^{\cal M}(\cos\theta^*)
        \left[
                P^+_{\cal J}c_{\cal M} W^{TT(+)}_{\cal JM}
                +P^-_{\cal J}s_{\cal M}
                \widetilde{W}^{TT(-)}_{\cal JM}
        \right]
        \label{68b}\\
\widetilde{W}^{TT} &=&
         4\pi\sum_{\cal JM}
        f^i_{\cal J}P_{\cal J}^{\cal M}(\cos\theta^*)
        \left[
                P^+_{\cal J}s_{\cal M}
                \widetilde{W}^{TT(+)}_{\cal JM}
                -P^-_{\cal J}c_{\cal M}
                W^{TT(-)}_{\cal JM}
        \right]
        \label{69b}\\
\widetilde{W}^{T'} &=& 4\pi\sum_{\cal JM}
        f^i_{\cal J}P_{\cal J}^{\cal M}(\cos\theta^*)
        \left[
                P^-_{\cal J}c_{\cal M} W^{T'(-)}_{\cal JM}
                +P^+_{\cal J}s_{\cal M}
                \widetilde{W}^{T'(+)}_{\cal JM}
        \right]
        \label{70b}\\
W^{TL'} &=&
         4\pi\sum_{\cal JM}
        f^i_{\cal J}P_{\cal J}^{\cal M}(\cos\theta^*)
        \left[
                P^-_{\cal J}s_{\cal M}
                \widetilde{W}^{TL'(-)}_{\cal JM}
                -P^+_{\cal J}c_{\cal M}
                W^{TL'(+)}_{\cal JM}
        \right]
        \label{71b}\\
\widetilde{W}^{TL'} &=& 4\pi\sum_{\cal JM}
        f^i_{\cal J}P_{\cal J}^{\cal M}(\cos\theta^*)
        \left[
                P^-_{\cal J}c_{\cal M} W^{TL'(-)}_{\cal JM}
                +P^+_{\cal J}s_{\cal M}
                \widetilde{W}^{TL'(+)}_{\cal JM}
        \right],
        \label{72b}
\end{eqnarray}
where we define $s_{\cal M}\equiv \sin({\cal M}\Delta\phi)$ and 
$c_{\cal M}\equiv \cos({\cal M}\Delta\phi)$, the factors that contain
all of the dependence on $\Delta\phi$, and
where for brevity we have only written the ranges of the sums over
$\cal J, M$ in the $L$ and $T$ pieces. In all of the others a sum 
for $\Jb=0,\ldots,2J_i$ and for
non-negative values of $\cal M$ must be understood. 
The $W$'s and $\widetilde{W}$'s inside the sums are the {\em angular
reduced response functions}.
Note that for the 
limit values $\theta^*=0^\circ,180^\circ$ the Legendre function
$P^{\cal M}_{\cal J}(\cos\theta^*)$ is zero for ${\cal M}\ne 0$.
Therefore the functions $W^{K}$, $\widetilde{W}^{K}$ do not depend on 
$\Delta\phi$ for these values of $\theta^*$. Note also
that the angular reduced response functions
$W^{K(\pm)}_{\cal JM}(q,\omega,\theta')$ are linear combinations of the 
reduced responses $W^{K(\pm)}_{\Jb\Jb'L}$ with the functions
$h_{\Jb\Jb'LM}^{\cal M}$ as coefficients, while the angular reduced responses 
with a tilde
$\widetilde{W}^{K(\pm)}_{\Jb\Jb'L}$ have as coefficients the functions
$\widetilde{h}_{\Jb\Jb'LM}^{\cal M}$. 
Furthermore, in Eqs.~(\ref{64b}--\ref{72b})
the responses without tildes are multiplied by 
$\cos({\cal M}\Delta\phi)$,
while the responses with tildes are multiplied by 
$\sin({\cal M}\Delta\phi)$.
They can be written as
\begin{eqnarray} 
        W^{K(\pm)}_{\cal JM}(q,\omega,\theta')
&=&     \sum_{\Jb' L}a_{\Jb' L} 
        h_{\Jb\Jb'LM}^{\cal M}(\theta')
        W^{K(\pm)}_{\Jb\Jb' LM}(q,\omega)
        \label{wjm}\\
        \widetilde{W}^{K(\pm)}_{\cal JM}(q,\omega,\theta')
&=&     \sum_{\Jb' L}a_{\Jb' L} 
        \widetilde{h}_{\Jb\Jb'LM}^{\cal M}(\theta')
        W^{K(\pm)}_{\Jb\Jb' LM}(q,\omega)
        \label{wjm-tilde},
\end{eqnarray}
where the index $M$ depends on the particular response as
\begin{equation}
M = \left\{ \begin{array}{cl}
                0 & \mbox{for $K=L,T,T'$} \\
                1 & \mbox{for $K=TL,TL'$} \\
                2 & \mbox{for $K=TT$,} 
            \end{array}
   \right.
\end{equation}
that is, the value of $M$ depends on the particular response 
in the same way as in Eqs. (\ref{50},\ref{55}--\ref{60}).
Finally, we have introduced a factor 
$a_{\Jb'L}=P^+_{\Jb'+L}$
for $K=T$, 
$a_{\Jb'L}=P^-_{\Jb'+L}$
for $K=T'$, while
$a_{\Jb'L}=1$ for the other responses.

    From these  general expressions, one can easily specialize to some 
important cases, such as:
\begin{enumerate}
\item {\em Nucleon knockout from spin-zero nuclei}. In this case
in the above equations we  have  $J_i=0$, 
and thus only the terms $\Jb=0$, $\Jb'=L$, $J_f=J$ and $J'_f=J'$
contribute. For brevity we do not give the explicit expressions for 
this case, as in this work we are considering only odd nuclei with
half-integer
spin.

\item {\em Inclusive responses}. Integrating over the emission angles
$\hp'$ and summing over the final nuclear states $|B\rangle$ (including 
over proton and neutron emission) we obtain 
the inclusive responses. Due to the presence of the spherical harmonic 
$Y_{\cal J'M'}(\hp')$, the integral over $\hp'$ in all of
the above equations restricts the values to $\Jb'=0$, $\Jb=L$, 
$j=j'$, $J_f=J'_f$, $\Mb=M$ and $l=l'$, yielding an alternative derivation
of
the equations given in Ref.~\cite{Ama97}
and also providing a test of the multipole expansion.

\item {\em Unpolarized responses for any spin $J_i$ nucleus.} 
The unpolarized responses can be computed as the
polarized responses averaged over all polarization
directions, that is
\begin{equation}\label{73b}
\Rb_{unpol}^K=\frac1{4\pi}\int \Rb^K(\Omega^*)\,d\Omega^*,
\end{equation}
and therefore only the terms with $\Jb=0$ in the above equations survive. 
These terms are obviously independent of $\Omega^*$ as is required for an
unpolarized observable, so the integral trivially yields a factor $4\pi$
that cancels with the denominator in Eq.~(\ref{73b}). 
As a consequence, the unpolarized responses are given by the 
terms with $\Jb=\Mb=0$ in Eqs.~(\ref{64b}--\ref{72b}).

\end{enumerate}

%========================================================

\section{Formalism for one-hole shell model nuclei }

%========================================================

As a particular application of the formalism introduced in the last section
to a simple model of a polarized target, in this work we consider
the case of a one-hole shell model nucleus. Specifically we
consider the same model introduced in Ref.~\cite{Ama97} for the
analysis of inclusive polarized responses and refer the reader to 
the discussions presented there for more details on the model. 

The initial nuclear state is a hole in a closed-shell core $|C\rangle$:
\begin{equation}
|A\rangle=|i^{-1}(\Omega^*)\rangle=
\sum_{m_i}{\cal D}^{(j_i)}_{m_ij_i}(\Omega^*)b_{i,m_i}^{\dagger}|C\rangle,
\end{equation}
where $|i\rangle=|n_i,\frac12,l_i,j_i\rangle$ is a single-particle
state occupied in the core and $b_{i,m_i}^{\dagger}$ is the creation
operator for a hole. Here we follow the convention of using lower case
letters 
$j_i$ for half-integer  angular momenta. As in the present work we only 
consider the one-body piece of the electromagnetic nuclear current, the
interaction with the virtual photon gives rise only to particle-hole
(p-h) excitations. Thus the final nuclear states are described by
\begin{equation}
|f\rangle=|(E'lj)(h^{-1},i^{-1})J_B;j_f\rangle.
\end{equation}
Here $|h\rangle=|n_h,l_h,j_h\rangle$ is another (bound) single-particle 
state in the core, while $|E'lj\rangle$ is a single particle in the
continuum.
The residual nucleus is represented in this simple model as a two-hole
nucleus
$|B\rangle=|(h^{-1},i^{-1})J_B\rangle$ with total angular momentum $J_B$, 
coupled with the outgoing particle $|E'lj\rangle$ to a total angular 
momentum $j_f$.

In the present work the wave functions of the single-particle hole states
are 
obtained using a mean-field potential of Woods-Saxon type. More details on 
this aspect of the calculation, including the values of the potential 
parameters, can be found in Refs.~\cite{Ama96} and \cite{Ama94}. On the
other 
hand, the wave function for the ejected nucleon is obtained as the solution
of 
the Schr\"odinger equation for positive energy $\epsilon'=E'-M$ ($E'$
contains
the rest mass), using a complex optical potential in order to include 
absorption processes in the FSI that are not described by a real mean-field
potential. Below we give more details about this particular aspect of the 
modeling.

The basic quantities required are the reduced response functions in
Eqs.~(\ref{61}--\ref{72}). The sums in those equations run over 
$\sigma=(l,j,J,j_f)$ and $\sigma'=(l',j',J',j'_f)$, but as in the inclusive
case \cite{Ama97}, we are able to perform the sum over $j_f,j'_f$ 
analytically, simplifying the final expressions and permitting us to draw 
several conclusions about the responses. We summarize
the 
procedure to be followed in Appendix B. Worth mentioning here is the fact
that 
in Ref.~\cite{Ama97} where we considered only the inclusive case we were
able 
to perform analytically an additional sum over the daughter angular
momentum  
$J_B$, considerably simplifying the procedures: here we cannot perform this
additional summation since the value of $J_B$ is fixed. 

It is noteworthy that in this model of coincidence scattering, after 
performing the sums that are permitted, the resulting nuclear responses
can be expressed in terms of the polarized response functions for a single 
particle in the shell $h$.
These single-particle responses are due to
the scattering of the electron with a single nucleon in the shell $h$,
and in this case Eq.~(\ref{35}) is also valid, with the exception that 
$J_i=j_h$ and $J_B=0$, as explained in Appendix B.
We denote the single-particle reponses of the shell $h$ by
$w^{K(\pm)}_{\cal JM}[h\rightarrow 0]$. First we consider
the case $h\ne i$. Then, as shown in Appendix B, the corresponding response
for the entire nucleus, that we denote by 
$W^{K(\pm)}_{\cal JM}\left[i^{-1}\rightarrow (h^{-1}i^{-1})J_B\right]$,
is proportional to the response of a single nucleon in the shell $h$ via
the
simple relationship
\begin{equation}\label{77b}
W^{K(\pm)}_{\cal JM}\left[i^{-1}\rightarrow (h^{-1}i^{-1})J_B\right]
= 
(-1)^{j_h+j_i+J_B+\Jb}[J_B]^2
\seisj{j_i}{j_i}{\cal J}{j_h}{j_h}{J_B}
w^{K(\pm)}_{\cal JM}[h\rightarrow 0]. 
\end{equation}
Actually in Appendix B we prove the result at the more elementary
level, namely between the reduced responses
$W^{K(\pm)}_{\Jb\Jb'L}\left[i^{-1}\rightarrow (h^{-1}i^{-1})J_B\right]$
and $w^{K(\pm)}_{\Jb\Jb'L}[h\rightarrow 0]$.

This relationship 
has important consequences. Indeed, although in this 
model the polarization of the nucleus is carried by the outer shell 
$i$, it is interesting to note that nucleon knockout can provide information 
about the polarized responses of the inner, complete (unpolarized) shell 
$h$, as long as the 6-$j$ 
coefficient is nonzero. In particular, if $i$ is a neutron (or a proton)
we can measure polarization observables with $(e,e'p)$ (or $(e,e'n)$) 
reactions. The reason for that behaviour is that the final particles 
are partially polarized because they are coupled to a definite 
angular momentum $J_B$.

For further insight into this last comment, let us consider the sum 
over all possible values of $J_B$
in Eq.~(\ref{77b}). Then we obtain the sum rule for the 
angular reduced responses
\begin{equation}
\sum_{J_B}
W^{K(\pm)}_{\cal JM}\left[i^{-1}\rightarrow (h^{-1}i^{-1})J_B\right]
= \delta_{\Jb 0}
  \delta_{\Mb 0}
  [j_i][j_h]
   w^{K(\pm)}_{00}[h\rightarrow 0], 
\end{equation}
and a similar result for the total responses. For instance, using 
Eq.~(\ref{64b}) for the longitudinal case we have
\begin{equation} \label{97}
\sum_{J_B}
\Rb^L\left[i^{-1}\rightarrow (h^{-1}i^{-1})J_B\right]
= 4\pi
  [j_h]w^{L(+)}_{00}[h\rightarrow 0], 
\end{equation}
where we have used $f^i_{0}=1/[j_i]$. We can compare this equation with the
longitudinal response for electron scattering from an unpolarized particle
in 
shell $h$. From Eq.~(\ref{73b}) and the comments that follow it, we see
that
the unpolarized response is given by the term with $\Jb=0$ in
Eq.~(\ref{64b}), 
with $J_i=j_h$
\begin{equation}
\Rb^L_{unpol}[h\rightarrow 0] 
= 4\pi\frac{1}{[j_h]}
  w^{L(+)}_{00}[h\rightarrow 0].
\end{equation}
Then, inserting this equation in Eq.~(\ref{97}), we obtain a relation that
is 
valid for all of the responses, $K=L,T,\ldots$, and not only for the 
longitudinal case:
\begin{equation} 
\sum_{J_B}\label{83}
\Rb^K\left[i^{-1}\rightarrow (h^{-1}i^{-1})J_B\right]
= [j_h]^2\Rb^{K}_{unpol}[h\rightarrow 0]. 
\end{equation}
This shows that the sum over $J_B$ of all of the exclusive
polarized $K$-responses for different final states of the type
$(h^{-1}i^{-1})J_B$ is equal to the unpolarized $K$-response
of the shell $h$ (that is, arising from unpolarized
nucleon knockout from a nucleon in the shell $h$) times the number
of particles in the complete shell $h$. As a consequence,
the sum over $J_B$ of all of the polarized responses (i.e. with $\Jb\ne0$)
is
equal to zero. Of course this theorem is only valid in the extreme shell
model, which means that observable deviations of this result
can provide valuable information about configuration mixing and 
the nuclear structure of the final states.
In summary, we see that the reason why unpolarized shells yield
polarization observables is due to the fact that the angular
momentum $J_B$ of the daughter nucleus is fixed. If we sum over
$J_B$ all of the polarization observables go away for a closed shell.

Now we consider the case $h=i$, where we extract a nucleon 
from the same shell in which the hole is located. This is the 
case with more practical applications, because
the shell model is expected to provide a better 
description of the nuclear states for small excitation energy
and, in particular, for the ground state of the daughter nucleus
with $J_B=0$. The derivation of the results that follows is
similar to the $h\ne i$ case, with the exception that a factor of 
two arises due to the fact that the two holes in the same 
shell are undistinguishable, and, in addition, that the nuclear
spin $J_B$ is restricted to be an even number.  Then, if we set
$h=i$ in Eq.~(\ref{77b}) and multiply by two, we obtain
\begin{equation} \label{84}
W^{K(\pm)}_{\cal JM}\left[i^{-1}\rightarrow (i^{-1}i^{-1})J_B\right]
= 
2(-1)^{\Jb+1}[J_B]^2
\seisj{j_i}{j_i}{\cal J}{j_i}{j_i}{J_B}
w^{K(\pm)}_{\cal JM}[i\rightarrow 0]. 
\end{equation}
In particular, for the daughter nucleus in the ground state
with $J_B=0$ the 6-$j$ symbol is equal to $(-1)^{\Jb+1}/[j_i]^2$ and we
have
 \begin{equation}
W^{K(\pm)}_{\cal JM}\left[i^{-1}\rightarrow (i^{-1}i^{-1})0\right]
= \frac{2}{[j_i]^2}
w^{K(\pm)}_{\cal JM}[i\rightarrow 0]. 
\end{equation}
For example,  for the simplest case
of  a hole in a $j_i=\frac12$ shell,
we have $[j_i]^2=2$ and therefore we obtain
 \begin{equation}
\textstyle
W^{K(\pm)}_{\cal JM}
\left[\frac12^{-1}\rightarrow (\frac12^{-1}\frac12^{-1})0\right]
=
w^{K(\pm)}_{\cal JM}\left[\frac12\rightarrow 0\right]. 
\end{equation}
Then we recover the result that one hole in a $\frac12$ shell is equivalent
to a particle in that shell, as expected. 

Also, if we sum over $J_B$ as before, we obtain an interesting sum rule. 
Taking into account that the sum now runs over $J_B=\mbox{even}$
and using the property 
\begin{equation}
\sum_{J_B=\mbox{\small even}} [J_B]^2
\seisj{j_i}{j_i}{\cal J}{j_i}{j_i}{J_B}
= \frac12(1- \delta_{\Jb 0}[j_i]^2)
\end{equation}
then, from Eq.~(\ref{84}),
 we obtain for the sum of the angular reduced responses 
\begin{equation}\label{88}
\sum_{J_B}
W^{K(\pm)}_{\cal JM}\left[i^{-1}\rightarrow (i^{-1}i^{-1})J_B\right]
=  (\delta_{\Jb 0}[j_i]^2-(-1)^{\Jb})
   w^{K(\pm)}_{\cal JM}[i\rightarrow 0], 
\end{equation}
i.e., the unpolarized response of the $2j_i+1$ particles
in the complete shell minus the polarized response of 
a single particle multiplied by $(-1)^{\Jb}$. 
The same relation holds between the sum 
$\sum_{J_B}
W^{K(\pm)}_{\Jb\Jb'L}\left[i^{-1}\rightarrow (i^{-1}i^{-1})J_B\right]$
and the responses
$w^{K(\pm)}_{\Jb\Jb'L}[i\rightarrow 0]$. 
The origin of the phase 
$(-1)^{\Jb}$ is easily explained by looking at the total responses. 
For instance, from Eq.~(\ref{55}) we obtain for the 
longitudinal one 
\begin{eqnarray}
\sum_{J_B}\Rb^L\left[i^{-1}\rightarrow (i^{-1}i^{-1})J_B\right]
        &=& [j_i]^2\Rb^L_{unpol}[i\rightarrow0] 
                \nonumber\\
        &&\kern -5cm
         -4\pi\sum_{\Jb\Jb'L}f^i_{\Jb}(-1)^{\Jb}
        \left\{ P^+_{\Jb}A_{\Jb\Jb'L0}
                w^{L(+)}_{\Jb\Jb'L}[i\rightarrow0]
             -  P^-_{\Jb}B_{\Jb\Jb'L0}
                w^{L(-)}_{\Jb\Jb'L}[i\rightarrow0]
        \right\}.
\end{eqnarray}
We notice that we can include the factor $(-1)^{\Jb}$ in the angular
 functions $A_{\Jb\Jb'L0}$ and $B_{\Jb\Jb'L0}$ using the 
parity property of the spherical harmonics 
$Y_{\cal JM}(-\nOmega^*)=(-1)^{\Jb}
Y_{\cal JM}(\nOmega^*)$. Then we can write the above result
for any response, and not only for the longitudinal case, as
\begin{equation}\label{90}
\sum_{J_B}\Rb^K\left[i^{-1}\rightarrow (i^{-1}i^{-1})J_B\right]
        =  [j_i]^2\Rb^K_{unpol}[i\rightarrow0] 
        - \left. \Rb^K[i\rightarrow0]
         \right|_{\nOmega^*\rightarrow-\nOmega^*}. 
\end{equation}
Now it becomes evident that the factor $(-1)^{\Jb+1}$
for the difference between particle and hole cases which  was
also mentioned in Ref.~\cite{Ama97} comes from the fact that 
a hole nucleus which is polarized in the direction $\nOmega^*$ is 
physically equivalent to the absence of a particle 
which is polarized in the opposite  direction $-\nOmega^*$.
The factor $(-1)^{\Jb}$ comes from a spherical harmonic
$Y_{\cal JM}(\Omega^*)$ which is always in front of the
responses of rank $\Jb$ in the multipole expansion. 

A similar result to Eq.~(\ref{90}) was obtained also in the 
inclusive case (see Ref.~\cite{Ama97}), where it was shown
that any polarized (inclusive) response of a hole 
nucleus can be written as the unpolarized response of the 
complete shell minus the polarized response of the hole
(multiplied again by the factor $(-1)^{\cal J}$). 
We can re-obtain here that result by integration of
Eq.~(\ref{90}), thereby showing another connection between
the two formalisms. 

To finish this section we show a result for the sum of
the unpolarized (exclusive) responses of a hole nucleus:
\begin{equation}
\sum_{J_B}
\Rb^K_{unpol}\left[i^{-1}\rightarrow (i^{-1}i^{-1})J_B\right]
        =  ([j_i]^2-1)\Rb^K_{unpol}[i\rightarrow0] ,
\end{equation}
i.e., the sum of the unpolarized responses for each
one of the particles in the shell. 

Equations (\ref{77b}) and (\ref{84}) are the basic 
results of this section. They show
that the reduced polarized responses corresponding to different
values of the final angular momenta of the daughter nucleus
$|(h^{-1}i^{-1})J_B\rangle$ are scaled by a well-defined 
coupling coefficient that only depends on the values of the 
angular momenta involved. This fact could be experimentally 
checked and any deviation from that relationship would imply interesting
conclusions about the nuclear structure of the daughter states.

%========================================

        \section{PWIA}

%========================================

%-------------------------------------------
\subsection{Factorization of the responses}
%-------------------------------------------

The main reason for considering the PWIA for the ejectile
is that in such a limit, when the final-state interaction is turned off, 
the exclusive cross section factorizes and we can perform part of the 
calculations analytically. The resulting cross section
contains the basic electron-nucleon (in general off-shell)
responses and the scalar and vector momentum distributions,
giving a picture of the physics that is more transparent than the
general formalism where the multipole analysis obscures 
the physical interpretation of the process.

In order to extract valuable information about the nuclear momentum 
distribution, it would be desirable to find a set of polarization
observables
that is more or less independent of the FSI. The general formalism 
for PWIA in polarized nuclei was developed in Refs.~\cite{Cab93,Cab94},
and it was recently specialized in Ref.~\cite{Ama97} to the particular case
of 
polarized, one-hole nuclei, although in that work the exclusive responses
were 
integrated over the emission angle in order to study the inclusive case. 
Here we briefly summarize the essential results from that work so that the 
nomenclature can be introduced, referring to Ref.~\cite{Ama97} 
for more detailed discussions.

In PWIA we write the factorized exclusive responses as 
\begin{eqnarray}
{\cal R}^K 
& = & Mp'\, w^K_SM^S(\np) \label{factorI}\\
{\cal R}^{K'} 
& = & Mp'\,\nw^{K'}_V\cdot\nM^V(\np), \label{factorII}
\end{eqnarray}
where $\np\equiv \np'-\nq$ is the missing momentum with angles 
$(\theta,\phi)$, and $w^{K}_S$ and $\nw^{K'}_V$ are respectively the 
scalar and vector single-nucleon responses, which are related 
with the components
Eqs.~(\ref{22}--\ref{27}),
of the single-nucleon tensor spin matrix 
\begin{equation}
w^{\mu\nu}(\np',\np) = 
    J^{\mu}(\np',\np)^{\dagger}
           J^{\nu}(\np',\np),
\end{equation}
where $ J^{\mu}(\np',\np)$ is the electromagnetic current 
spin matrix for the nucleon, evaluated with plane waves 
$\langle \nr|\np s\rangle = (2\pi)^{-3/2}{\rm e}^{i\np\cdot\nr}\chi_s$.
Taking the appropriate components of  the single-nucleon tensor 
$w^{\mu\nu}(\np',\np)$, the resulting unprimed responses 
are diagonal in spin (spin-scalar), so we can write for the 
$L$, $T$, $TL$ and $TT$ responses
$ w^K= w^K_S$, 
while the $T'$ and $TL'$ (primed) responses are spin-vectors and can 
be written as
$w^{K'}=\nw^{K'}_V\cdot\nsigma$.
Explicit expressions for the current matrices and single-nucleon
responses are given in Ref.~\cite{Ama97}.

The momentum distribution for residual nucleus $|B\rangle$ is defined
in general as the spin matrix 
\begin{equation} \label{98}
n(\np)_{s's} = \sum_{M_B}
\langle B | a_{\np,s}|A\rangle^*
\langle B | a_{\np,s'}|A\rangle,
\end{equation}
where the spin $J_B$ is fixed and we sum over undetected orientations
of the daughter nucleus.
Therefore the definition of the scalar $M^S$ and vector $\nM^V$ 
momentum distributions is 
\begin{equation}\label{99}
n(\np) = \frac12\left(M^S(\np)+\nM^V(\np)\cdot\nsigma\right).
\end{equation}
Obviously, the polarized momentum distribution depends upon the initial
state 
$|A\rangle$, the final state $|B\rangle$ and the polarization angles 
$\Omega^*$, apart from the missing momentum $\np$. When we need to clarify 
this dependence below, we sometimes write in the momentum distribution 
the explicit process to which we are referring as an argument, for example,
$n\left[ A(\Omega^*)\rightarrow B\right]$.

%--------------------------------------------------
\subsection{Momentum distributions of hole nuclei}
%--------------------------------------------------

Here we obtain the polarized momentum distribution 
in the particular case of a hole nucleus. For the general expression
of the polarized spectral function of deformed nuclei, see
Ref.~\cite{Cab94}. 
In this work we are also interested in examining
in more detail the relations between the single-particle
momentum distribution and that of the whole nucleus in the extreme
shell model, and thus we will follow a somewhat different approach
(although one that is completely equivalent). 
For this reason we begin examining the polarized momentum
distribution of a single particle (or hole) in a shell.

\subsubsection{Single-particle momentum distribution}

Let us consider a single particle in the bound state
$|jj(\Omega^*)\rangle=\sum_m{\cal D}^{(j)}_{mj}|jm\rangle$
of a mean field; here one should understand there to be 
the argument $\Omega^*$ 
in all the rotation matrices, polarized states and momentum distributions. 
In this simple case, the momentum distribution can be written 
as the matrix element
\begin{equation}
n_{s's}[j(\Omega^*)]=
\langle\np s'| N[j(\Omega^*)] | \np s \rangle
\end{equation}
of the spin density operator $N[j(\Omega^*)]$ which projects onto the 
state $|jj*\rangle$
\begin{equation}
N[j(\Omega^*)]  
        = |jj*\rangle\langle jj*| 
        = \sum_{mm'}
          {\cal D}^{(j)}_{m'j}{\cal D}^{(j)*}_{mj}
          |jm'\rangle\langle jm|.
\end{equation}
Expanding  the product of the two rotation
matrices, we are able to write a multipole expansion 
for the operator $N[j(\Omega^*)]$ in a basis of spherical 
harmonics of the polarization angle:
\begin{equation}
N[j(\Omega^*)]  
= \sum_{\cal J} f^j_{\cal J}
N^{\cal J}[j(\Omega^*)],  
\end{equation}
where $N^{\cal J}[j(\Omega^*)]$ is the operator of rank $\cal J$ in the 
expansion, given by
\begin{equation}\label{104}
N^{\cal J}[j(\Omega^*)]  
= \sqrt{4\pi}
  \sum_{mm'\cal M} 
  (-1)^{j+m+\cal J}
  \tresj{j}{j}{\cal J}{-m}{m'}{\cal M}
  Y_{\cal JM}(\Omega^*)
  |jm'\rangle\langle jm|.
\end{equation}  

A similar expansion can be written for 
the case of a single hole, namely, the focus 
in the present work. In the shell model a hole is described by
 a single-particle state with wave
function 
$|\widetilde{jj*}\rangle= \sum_m {\cal D}^{(j)*}_{mj}|\widetilde{jm}\rangle$,
 i.e., the time inversion of
the single-particle state, with 
$|\widetilde{jm}\rangle=(-1)^{j+m}|j,\mbox{$-m$}\rangle$. 
In this case one can prove that 
the corresponding
spin-density matrix operator multipole expansion can be written as
\begin{equation}
N[\widetilde{j(\Omega^*)}]  
= \sum_{\cal J} f^j_{\cal J}
(-1)^{\cal J}
N^{\cal J}[{j(\Omega^*)}],  
\end{equation}
were we have used the fact that the  spin-density multipoles  
of particles and holes differ only by the phase factor 
$(-1)^{\cal J}$.
This a consequence of the fact that the state 
$|j,-j\rangle$ rotated by an angle $\Omega^*$ is the same as the state 
$|jj\rangle$ rotated by an angle $-\Omega^*$ and of the parity 
$(-1)^{\cal J}$  of the spherical harmonics which are the basis of
 the multipole expansion.

\subsubsection{ Hole nucleus}

In analogy with the single-particle case, we can write the
momentum distribution as the matrix element of a one-body 
density matrix. In fact, consider the transition
\begin{equation}
|A\rangle=|i^{-1}(\Omega^*)\rangle\rightarrow|B\rangle
= |(h^{-1}i^{-1})J_B\rangle.
\end{equation}
Inserting the appropriate matrix elements of the destruction operator
into the momentum distribution, Eq.~(\ref{98}), we obtain
\begin{equation}
n_{s's}(\np)=
\langle \np s'| N\left[i^{-1}(\Omega^*)\rightarrow(h^{-1}i^{-1})J_B\right]
|\np s\rangle,
\end{equation}
where $N\left[i^{-1}(\Omega^*)\rightarrow(h^{-1}i^{-1})J_B\right]$
is the spin density matrix operator
\begin{eqnarray}
N\left[i^{-1}(\Omega^*)\rightarrow(h^{-1}i^{-1})J_B\right]
&=& \nonumber\\
&&\kern -5cm 
\sum_{M_B}\sum_{m_hm'_i}\sum_{m'_hm''_i}
\langle j_hm_h j_im'_i| J_BM_B\rangle
\langle j_hm'_h j_im''_i| J_BM_B\rangle
{\cal D}^{(j_i)*}_{m'_ij_i}{\cal D}^{(j_i)}_{m''_ij_i}
|\widetilde{hm'_h}\rangle\langle\widetilde{hm_h}|.
\end{eqnarray}
Expanding again  the product of the two rotation matrices
 we obtain a sum of the product of three 3-$j$ coefficients,
that can be written as a 6-$j$ coefficient times a 3-$j$
coefficient. As consequence, we straightforwardly obtain
the multipole expansion 
\begin{equation}
N\left[i^{-1}(\Omega^*)\rightarrow(h^{-1}i^{-1})J_B\right]
=
\sum_{\cal J}f^i_{\cal J}
N^{\cal J}\left[i^{-1}(\Omega^*)\rightarrow(h^{-1}i^{-1})J_B\right],
\end{equation}
with the simple relationship between the nuclear multipoles
and the single-hole $h$ results:
\begin{equation} \label{116}
N^{\cal J}\left[i^{-1}(\Omega^*)\rightarrow(h^{-1}i^{-1})J_B\right]
= [J_B]^2(-1)^{j_i+j_h+J_B+\cal J}
\seisj{j_h}{j_h}{\cal J}{j_i}{j_i}{J_B}
N^{\cal J}[h(\Omega^*)].
\end{equation}
This relation is the same as the one found between the
reduced nuclear responses and the responses of the shell $h$
(see Eq.~(\ref{77b})). In fact, the general
equations of the last section are of course also
applicable to the PWIA when we take the 
single-particle state $|E'lj\rangle$ to be a free wave function,
i.e, proportional to a  Bessel function. Therefore, Eq.~(\ref{77b})
had to be valid also in PWIA. And that fact is again recovered here in
another completely different way, providing a test of both formalisms. 
But also the fact that the relation between momentum distributions
can be extended to response functions even with FSI leads us to
believe that some information about the single-particle
momentum distribution can be extracted from these kinds
of observables. It is remarkable that, although the FSI
involving the ejected nucleon destroys the factorization of the
cross section, it does not modify the relation in Eq.~(\ref{116}),
implying that this property is general for the independent-particle model. 

Until now we have only considered the case $h\ne i$. The 
other situation $h=i$ that arises when the particle is ejected
from the polarized shell is analogous, with the exception that 
now the residual nucleus has a wave function with a factor
$2^{-1/2}$
\begin{equation}
|B\rangle = 
\frac{1}{\sqrt{2}}[b_i^{\dagger}b_i^{\dagger}]_{J_BM_B}|C\rangle.
\end{equation}
Therefore the expression for the momentum distribution as the 
matrix element of a spin density matrix operator is 
\begin{equation}
n_{s's}\left[i^{-1}(\Omega^*)\rightarrow(i^{-1}i^{-1})B\right]
=\langle \np s'|
N\left[i^{-1}(\Omega^*)\rightarrow(i^{-1}i^{-1})B\right]
|\np s\rangle
\end{equation}
with 
\begin{equation} \label{121}
N^{\cal J}\left[i^{-1}(\Omega^*)\rightarrow(i^{-1}i^{-1})J_B\right]
= 2[J_B]^2(-1)^{1+\cal J}
\seisj{j_i}{j_i}{\cal J}{j_i}{j_i}{J_B}
N^{\cal J}[i(\Omega^*)],
\end{equation}
and we again recover the relation in Eq.~(\ref{84}) between reduced
response functions in the shell model. 

These relations between the multipole components of the 
density matrix lead us to obtain sum rules when summing over the 
the values of $J_B$, as we did for the responses in the shell model. 
In fact, performing the sum over $J_B$ in the
case $h\ne i$ we obtain
a the result analogous to Eq.~(\ref{83})
\begin{equation}\label{124}
\sum_{J_B}
N\left[i^{-1}(\Omega^*)\rightarrow(h^{-1}i^{-1})J_B\right]
=[j_h]^2 N[h]_{unpol},
\end{equation}
where
$N[h]_{unpol}$ in the 
umpolarized density matrix of the single-particle wave function.
This result was also obtained in Ref.~\cite{Ama97} in relation
with the inclusive PWIA, where the final state $B$ is
not observed and the sum over $J_B$ must be performed. 
Strictly speaking, what in Ref.~\cite{Ama97} is called
{\em polarized momentum distribution

of a shell}  actually is the matrix element of
the sum over $J_B$ given by Eq.~(\ref{124}). Then we
can say that the momentum distribution
of a complete shell ($h\ne i$) is the sum over all of the 
unpolarized momentum distributions of all of the particles
in the shell. 

In the case $h=i$ we have the restriction $J_B=\mbox{even}$. 
The sum over $J_B$ gives an equation similar to Eq.~(\ref{88}),
and the total density matrix is 
\begin{equation}
\sum_{J_B=\rm even}
N\left[i^{-1}(\Omega^*)\rightarrow(h^{-1}i^{-1})J_B\right]
=[j_i]^2N[i]_{unpol}
-N[i(-\Omega^*)],
\end{equation}
which corresponds to Eq.~(\ref{90}) at the response level.
This result was also obtained in Ref.~\cite{Ama97} for 
the polarized momentum distribution of a shell with a hole:
it is equal to the unpolarized momentum distribution of the
complete shell minus the polarized momentum distribution
of the hole, although in this work we are able
to explain geometrically the origin of the factors $(-1)^{\cal J}$ that
appeared in that reference and also to understand the fact that
the momentum distribution of a hole is equal to the
momentum distribution of a particle pointing in the 
opposite direction. 

%------------------------------------------------------
\subsection{Scalar and vector momentum distributions}
%------------------------------------------------------

Until now we have seen that when the polarized momentum
distribution (or in general the spin density matrix operator)
of hole nuclei is expanded in multipoles of the polarization
angles, each one of the individual multipoles is proportional 
to the corresponding multipole of the single-particle
momentum distribution of the struck nucleon. Therefore
the same proportionality relation also holds for the scalar 
and vector pieces in Eq.~(\ref{99}) of that spin operator.
Thus we only need to describe the scalar and vector momentum
distributions of a single particle in the state $|j(\Omega^*)\rangle$. 
A detailed derivation of these expressions was performed 
in Ref.~\cite{Ama97}. Therefore here we only write the
results in a way more suited to use in the 
exclusive responses. The multipole expansion for the
scalar and vector momentum distributions reads
\begin{eqnarray}
M^S[j(\Omega^*)] & = & \sum_{\cal J}f^j_{\cal J}P^+_{\cal J}
                        M^S_{\cal J}[j(\Omega^*)] 
                        \label{multipoleMS}\\
\nM^V[j(\Omega^*)] & = & \sum_{\cal J}f^j_{\cal J}P^-_{\cal J}
                        \nM^V_{\cal J}[j(\Omega^*)], 
\end{eqnarray}
where in addition we have used the result that only
even (odd) multipoles enter in the expansion of the
scalar (vector) momentum distribution for a single particle.
Although not explicitly written, the functions
$M^S_{\cal J}$ and $\nM^V_{\cal J}$ depend on $(\Omega^*,\np)$. 
The only scalar that can be constructed with $Y_{\cal JM}(\Omega^*)$ 
and $\np$ is $[Y_{\cal J}(\Omega^*)Y_{\cal J}(\hp)]_{00}$,
therefore the scalar momentum distribution for a single particle
has the form
\begin{equation}\label{129}
M^S_{\cal J}(\Omega^*,\np) = m^S_{\cal J}(p)
[Y_{\cal J}(\Omega^*)Y_{\cal J}(\hp)]_{00}.
\end{equation}
Also, the vector momentum distribution can be written as a linear 
combination
\begin{equation}\label{130}
\nM^V_{\cal J}(\Omega^*,\np) = \sum_{\Jb'=\Jb\pm1}
m^V_{\Jb\Jb'}(p)
\nX_{\Jb\Jb'}(\Omega^*,\np)
\end{equation}
of the vectors 
$\nX_{\Jb\Jb'}(\Omega^*,\np)$
with spherical components
\begin{equation}
X_{\Jb\Jb'}(\Omega^*,\np)_{\alpha}\equiv
[Y_{\cal J}(\Omega^*)Y_{\cal J'}(\hp)]_{1\alpha}.
\end{equation}
Finally, the functions $m^S_{\cal J}(p)$ and 
$m^V_{\Jb\Jb'}(p)$ for a particle are given by
\begin{eqnarray}
m^S_{\cal J}[j(\Omega^*)] 
                &=& 2A_{\cal J}[j]^2|\widetilde{R}(p)|^2
                \label{147}\\
m^V_{\cal JJ'}[j(\Omega^*)] 
                &=& -\frac{2}{\sqrt{3}}
                    A_{\cal JJ'}[j]^2|\widetilde{R}(p)|^2,
\end{eqnarray}
where $\widetilde{R}(p)=\sqrt{2/\pi}\int_0^{\infty}{\rm d}r\, r^2 j_l(pr)R(r)$ 
is the radial wave function in momentum space, normalized to 
$\int_0^{\infty}{\rm d}p\, p^2 |\widetilde{R}(p)|^2=1$,
and the factors $A_{\cal J}$ and $A_{\cal JJ'}$ are coupling coefficients
which can be found in Ref.~\cite{Ama97}.
Finally, the same relations in Eqs.~(\ref{129},\ref{130}) are valid for
the scalar and vector momentum distributions 
for the process $i^{-1}(\Omega^*)\rightarrow(h^{-1}i^{-1})J_B$
and we have the following result:
\begin{eqnarray}
m^S_{\cal J}\left[i^{-1}(\Omega^*)\rightarrow(h^{-1}i^{-1})J_B\right]
&=& \nonumber\\
&& \kern -3cm
   (1+\delta_{hi})[J_B]^2(-1)^{j_i+j_h+J_B+\cal J}
   \seisj{j_h}{j_h}{\cal J}{j_i}{j_i}{J_B}
   m^S_{\cal J}[h(\Omega^*)]
   \label{137}
\end{eqnarray}
and a similar result for 
$m^V_{\cal JJ'}\left[i^{-1}(\Omega^*)\rightarrow(h^{-1}i^{-1})J_B\right]$
in terms of the single particle vector multipole
$m^V_{\cal JJ'}[h(\Omega^*)]$.

%------------------------------------------------
\subsection{Single-nucleon current and exclusive 
            nuclear responses in PWIA}
%------------------------------------------------

For the single-nucleon current we follow the formalism of 
Refs.~\cite{Ama96,Ama97}, where the on-shell relativistic
electromagnetic single-nucleon current was expanded up to order 
$\eta=p/M$, but keeping all orders and not expanding in
the dimensionless variables $\kappa=q/2M$ and $\lambda=\omega/2M$.
The expressions for the charge and transverse current components
in the coordinate system used in this work are given in 
Ref.~\cite{Ama97}, where also can be found 
the  expressions for the single-nucleon 
responses $w^K$. The important point here, and used for the 
results of next section, is that 
the $L$, $T$, $TL$ and $TT$ single-nucleon 
responses are spin-independent,
while the $T'$ and $TL'$ ones involve linear combinations of the 
Pauli spin matrices.

We now have all of the ingredients needed to obtain the reduced responses
of a one-hole nucleus in PWIA. The following equations have been derived
for a one-hole nucleus, although actually they are a little more general
---
in fact, they are valid for any nuclear model as long as the 
multipoles of the scalar and vector momentum distributions can be written 
as in Eqs.~(\ref{129},\ref{130}). 

For comparison with the general model it is 
convenient to write the responses in a form similar to 
Eqs.~(\ref{64b}--\ref{72b}). Therefore, for the unprimed responses that 
are proportional to the scalar momentum distribution $M^S$, 
we develop the (real) scalar coupling of two spherical 
harmonics as in Eq.~(\ref{76}):
\begin{equation}
[Y_{\cal J}(\Omega^*)Y_{\cal J}(\hp)]_{00} = 
  \sum_{\Mb=0}^{\cal J}
  P_{\cal J}^{\cal M}(\cos\theta^*)
  \cos[{\cal M}\Delta\phi]\, 
  h_{\Jb\Jb00}^{\cal M}(\theta),
\end{equation}
and so we have 
\begin{equation}
M^S_{\cal J}(\Omega^*,\hp)=\sum_{\Mb=0}^{\cal J}
P_{\cal J}^{\cal M}(\cos\theta^*)
\cos[{\cal M}\Delta\phi]\,
h_{\Jb\Jb00}^{\cal M}(\theta)
m^S_{\cal J}(p),
\label{sccal}
\end{equation}
where $m^S_{\cal J}$ refers to the considered transition
(see Eq.~(\ref{137})). For the vector momentum distribution we need to 
develop the three components of the vector $\nX_{\cal JJ'}$ in terms of the
functions $h_{\Jb\Jb'11}^{\cal M}$, $\widetilde{h}_{\Jb\Jb'11}^{\cal M}$
and $h_{\Jb\Jb'10}^{\cal M}$ in a similar way. 

As for the nuclear responses, using Eq.~(\ref{sccal}) we see that the scalar 
momentum distribution depends on the azimuthal angles only via
$\cos{\cal M}\Delta\phi$, where $\Delta\phi=\phi'-\phi^*$, but not 
on $\phi=\phi'$ or $\phi^*$ independently. The only dependence of 
the nuclear responses
on $\phi$ is therefore through the single-nucleon responses
$w^{TL}$ and $w^{TT}$, see Ref.~\cite{Ama97}. 
Then it is straightforward to read off
the following expressions for the angular reduced responses:
\begin{eqnarray}
W^{L(+)}_{\cal JM} &=& \frac{Mp'}{4\pi}
                        (\rho_c^2+\rho_{so}^2\delta^2)
                        h_{\Jb\Jb00}^{\cal M}(\theta)
                        m^S_{\cal J}(p)
                        \label{169}\\
W^{T(+)}_{\cal JM} &=& \frac{Mp'}{4\pi}
                        (2J_m^2+J_c^2\delta^2)
                        h_{\Jb\Jb00}^{\cal M}(\theta)
                        m^S_{\cal J}(p)
                        \\
W^{TL(+)}_{\cal JM} &=& \frac{Mp'}{4\pi}2\sqrt{2}
                        (\rho_cJ_c+\rho_{so}J_m)\delta
                        h_{\Jb\Jb00}^{\cal M}(\theta)
                        m^S_{\cal J}(p)
                        \label{171}
                        \\
W^{TT(+)}_{\cal JM} &=& -\frac{Mp'}{4\pi}
                        J_c^2\delta^2
                        h_{\Jb\Jb00}^{\cal M}(\theta)
                        m^S_{\cal J}(p).
                        \label{172}
\end{eqnarray}
The odd-rank unprimed responses and the
$\widetilde{W}_{\cal JM}^{TL(+)}$ 
and $\widetilde{W}_{\cal JM}^{TT(+)}$ 
responses are zero in PWIA. 
Here $\delta=\eta\sin\theta$ 
labels the order of the relativistic correction
and $(\theta,\phi)$ is the direction of the missing momentum $\np$. 
The factors $\rho_c$ (charge), $\rho_{so}$ (spin-orbit),
$J_m$ (magnetization) and $J_c$ (convection) come from the 
electromagnetic charge and current operators; they 
are only $(q,\omega)$-dependent, and are defined by (see Ref. \cite{Ama97})
\begin{eqnarray}
\rho_c    &=& \frac{\kappa}{\sqrt{\tau}}G_E \label{157}\\ 
\rho_{so} &=& \frac{2G_M-G_E}{\sqrt{1+\tau}}\frac{\kappa}{2}\label{158}\\
J_m       &=& \sqrt{\tau}G_M \label{159}\\
J_c       &=& \frac{\sqrt{\tau}}{\kappa}G_E, \label{160}
\end{eqnarray}
where $G_E$, $G_M$ are the electric and magnetic form factors of
the nucleon for which we use the Galster parametrization \cite{Galster}.

Finally, for the primed responses, after performing the scalar product
between $\nM^V$ and $\nw^{K'}_V$ we obtain
\begin{eqnarray}
W^{T'(-)}_{\cal JM} &=& -\frac{Mp'}{2\pi}
                        \sum_{\Jb'=\Jb\pm1}
                        \left[J_m^2
                              h_{\Jb\Jb'10}^{\cal M}(\theta)
                              + \sqrt{2}J_cJ_m\delta
                              h_{\Jb\Jb'11}^{\cal M}(\theta)
                        \right]
                        m^V_{\cal JJ'}(p)
                        \label{173}\\
W^{TL'(-)}_{\cal JM} &=& -\frac{Mp'}{\sqrt{2}\pi}
                        \sum_{\Jb'=\Jb\pm1}
                        \left[\sqrt{2}\rho_cJ_m
                              h_{\Jb\Jb'11}^{\cal M}(\theta)
                              +\rho_{so}J_m\delta
                              h_{\Jb\Jb'10}^{\cal M}(\theta)
                        \right]
                        m^V_{\cal JJ'}(p)
                        \\
\widetilde{W}^{TL'(-)}_{\cal JM} 
                &=& -\frac{Mp'}{\pi}
                        \sum_{\Jb'=\Jb\pm1}
                        (\rho_cJ_m-\rho_{so}J_c\delta^2)
                        \widetilde{h}_{\Jb\Jb'11}^{\cal M}(\theta)
                        m^V_{\cal JJ'}(p).
                \label{174}
\end{eqnarray}
All of the other responses not written here are exactly zero in PWIA.

%========================================================

 \section{Results and discussion}

%========================================================

%---------------------------------------
%\subsection{Details of the calculation}
%---------------------------------------

In this section we present numerical results for the reaction 
$^{39}\vec{\rm K}(\vec{e},e'p)^{38}$Ar in the shell model for both PWIA 
and with FSI, namely the distorted-wave impulse approximation (DWIA). 
An overview is provided in Sec. 5.1, followed by Sec. 5.2 
where detailed results are shown and discussed. 
Section 5.3 contains a brief discussion of the relativistic corrections
 incorporated in our model. Throughout all of these sections we have 
had to select only  the most important results for presentation,
since the entire set vastly exceeds what can be shown in published 
form. A more complete set of figures may be found on the WWW 
\cite{www97}. The reader is directed to the conclusions in the next section 
for discussion of what can be learned from such studies. 

The bound states of the target and daughter nuclei are represented by 
one-hole and two-holes in the core of $^{40}$Ca, as discussed in Sec.~2 
and in Ref.~\cite{Ama97}. It is clear that more sophisticated treatments 
of nuclei such as  $^{39}$K and $^{38}$Ar than the extreme shell model 
we are using in this work can also be studied, however, since our aims 
here and in the next paper~\cite{Ama98} 
are to analyze how FSI effects are reflected in the various 
polarization observables and to determine if it is possible to obtain 
information about the different pieces 
of the polarized momentum distribution or about the FSI with these kinds of
measurements, this basic model should provide an adequate starting point. 
Indeed, given the complexity that already exists for our basic model and 
the fact that no such extensive study for a complex nucleus is available,
the initial studies presented in Refs.~\cite{Cab94,Gar95,Bof88}
 being all that we  are aware of for coincidence 
polarization observables, at this time explorations with more
sophisticated nuclear configurations are probably unwarranted. Of course, 
when experimental studies of this type become imminent these extensions 
can be undertaken.

In our modeling the single-particle bound wave functions 
and (negative) energies are obtained using a mean-field
potential of Woods-Saxon type, using the values
of the potential parameters that may be found in Refs.~\cite{Ama94,Ama94b}.
For the ejectile wave function we solve the Schr\"odinger equation 
for positive energies using a (local) complex optical potential fitted to
proton elastic scattering data. For the particular
case of the nucleus $^{39}$K we use the potential 
of Schwandt {\em et al.} \cite{Sch82}, that was fitted 
to cross section and analyzing power angular
distributions for a number of target nuclei  
and polarized protons of 80--180 MeV. Although the
daughter nucleus $^{38}$Ar was not included in this
analysis, it falls in the range $24<A<208$  of applicability of
the Schwandt potential.  
As a first test of our model, we have computed
the cross section and analyzing powers for proton scattering 
from a variety of nuclei with different optical potentials
and have compared the results with  the literature. 
In particular, we have reproduced the 
calculations shown in Figs.~2 and 3 of 
Ref.~\cite{Sch82} for the analyzing power of 
protons elastically scattered from
$^{24}$Mg, $^{28}$Si, $^{40}$Ca, $^{90}$Zr and $^{208}$Pb
at different energies, using the Schwandt potential.

The normalization of the partial wave 
$(l,j)$ with energy $\epsilon$ and wave number
$k=\sqrt{2M\epsilon}$ is determined by the following asymptotic
condition for the radial wave function
\begin{equation}
R_{lj}(r) \sim 
\sqrt{\frac{2M}{\pi\hbar^2 k}}
{\rm e}^{i(\sigma_l+\delta_{lj})}
\sin\left( kr-\eta\log2kr-l\frac{\pi}{2}+\sigma_l+\delta_{lj}\right),
\end{equation}
where $\delta_{lj}$ is the complex nuclear phase-shift and $\sigma_l$ 
is the Coulomb phase-shift. 
In the limit when  the imaginary part of the potential goes to
zero, the radial wave functions are normalized with 
a Dirac $\delta$-function containing the energies,
so that condition in Eq.~(\ref{21}) is satisfied. 
The imaginary part of the potential modifies the normalization
of the continuum states, since the flux of outgoing
particles is reduced due to the absorptive property of
the optical potential. In addition, in this model 
the continuum nuclear wave
functions are no longer orthogonal to the bound ones.
However, at high momentum transfer 
$q\sim 500$~MeV/c 
and near quasi-free conditions where $p'\sim q$, the orthogonality between 
the bound and high-momentum ejected particles is approximately satisfied.
Several prescriptions to restore orthogonality
and to take into account non-locality effects have been
proposed, such as the inclusion of Perey-like factors \cite{Bof82}
in the wave functions. 
In our calculations  we include a Perey factor of the kind
\begin{equation}
f(r)=\left[1-\frac{M\beta^2}{2\hbar^2}{\rm Re}V_C(r)\right]^{-1/2},
\end{equation}
where $V_C$ is the central part of the optical potential and
we take the non-locality parameter $\beta=0.85$~fm.

In order to take into account relativistic
effects that are important for high momentum transfer
we use relativistic kinematics in all of the calculations \cite{Ama97}.
For nucleon knockout from shell $h$ 
we compute the momentum of the ejected particle 
with kinetic energy $\epsilon'=\epsilon_h+\omega$ 
using the relativistic energy-momentum relation 
${p'}^2=(M+\epsilon')^2-M^2$. In the shell model we solve the 
Schr\"odinger equation with eigenvalue 
$\epsilon'(1+\epsilon'/2M)$. 

In DWIA the sums over partial waves or multipoles $J$ 
are infinite, while in practice we must sum all multipoles
until convergence is reached. 
In this work we test the degree of convergence by 
comparing with the exact result in factorized PWIA.
In fact, in the limit in which the optical potential goes to zero, 
the DWIA and PWIA approaches should be identical and 
therefore in that limit we adjust the  number of multipoles 
so that differences with exact PWIA responses are indistinguishable. 
The number of multipoles needed to reach convergence is in 
general greater that in the inclusive case \cite{Ama97}. For example, for 
$^{39}$K at $q=500$ and 700~MeV/c in the region of the quasielastic peak
it is enough to include multipoles up to $J=15$ 
and 23, respectively, for the inclusive responses, while for the exclusive 
process we need to sum up to $J=22$ and 32, respectively. The reason is
that 
in the inclusive case the integration over
the angles $\hp'$ of the ejectile is performed analytically.
In all of the calculations that follow we include multipoles up to
$J=32$.

%--------------------------------------------
\subsection{Overview of the responses}
%--------------------------------------------

We apply the formalism of previous sections to the nucleus
$^{39}$K. In our model we describe the ground state as
a proton hole in the shell $1d_{3/2}$. Therefore we have
$J_i=3/2$. Since ${\cal J}=0,\ldots,2J_i$, the angular reduced response 
functions $W^{K(\pm)}_{\cal JM}$ 
and $\widetilde{W}^{K(\pm)}_{\cal JM}$ 
that enter in Eqs.~(\ref{64b},\ref{72b}) 
range from 
$({\cal J,M})=(0,0)$ to 
$({\cal J,M})=(3,3)$. 
Taking into account that all of the
responses with a tilde are zero for $\Mb=0$ 
(because they are defined as the coefficient of $\sin\Mb\Delta\phi$),
we are confronted with a total  of 72 angular reduced response functions 
(only 32 of them are nonzero in PWIA). 

In this work we focus our study of polarization observables
upon a systematic analysis of the complete set of angular
reduced response functions, which contain all of the information about the
process. An analysis of more direct 
observables, such as cross sections and asymmetries for different
values of the polarization angle is in progress \cite{Ama98}. 

We discuss only the case in which the daughter 
nucleus $^{38}$Ar is in the ground state $J_B=0^+$,
which we describe as a $d_{3/2}$ two-hole nucleus.
Under knockout from the $d_{3/2}$ shell the daughter
nucleus also can be in a $J_B=2^+$ state, but as
proven in Sec.~3, the reduced response functions
$W^{K(\pm)}_{\cal JM}$ for the two possible values 
of $J_B$ differ only in a constant recoupling coefficient
(see Eq.~(\ref{84})) and therefore they are proportional in
this simple model, so no new information is obtained from their 
analysis. Of course, in a more sophisticated treatment
of the nuclear structure, configuration mixing will produce 
deviations from this proportionality theorem, which may prove to be useful 
in obtaining new information about the nuclear structure involved.
 
In theory the angular reduced response functions $W^{K(\pm)}_{\cal JM}$ 
and $\widetilde{W}^{K(\pm)}_{\cal JM}$ could be extracted by performing a 
number of measurements with different electron helicity $h$, target 
polarization $\Omega^*$, azimuthal emission  angle $\phi$, and,
additionally, 
a Rosenbluth analysis in order to separate the longitudinal from the
transverse responses. The number of ways to carry out the extraction 
is not unique. For example, a discussion of the minimal asymmetry 
measurements needed to extract the reduced response functions in 
deuteron electrodisintegration is given in Ref.~\cite{Are92}, although 
in the $^{39}$K case the presence of the octupole 
terms ${\cal J}=3$, which do not appear in the deuteron case, modifies the 
extraction rules given in that reference. 
Rather than discussing the separation procedure or even to attempt to 
discuss the entire set of reduced responses in great detail, 
our goal in the following analysis in the present work is
to take into account that some of these reduced responses 
are very small relative to the unpolarized ones, and hence to focus on
those 
with the largest magnitude. We postpone any detailed discussion of 
cross sections and asymmetries for future presentation~\cite{Ama98}.

The angular reduced response functions are defined in 
Eqs.~(\ref{64b}--\ref{72b}). Those without a tilde 
$W^{K}_{\cal JM}$ 
are multiplied in the cross section by 
$\cos{\cal M}\Delta\phi$, 
while the responses with a tilde
$\widetilde{W}^{K}_{\cal JM}$ 
are multiplied by 
$\sin{\cal M}\Delta\phi$. 
Therefore the latter are zero for ${\cal M}=0$ or $\Delta\phi=0$, and hence
the tilde responses do not contribute to the cross section 
when the target is longitudinally polarized, i.e. in the direcction
of $\nq$ ($L$-polarized) or $-\nq$, because in those cases
$\cos\theta^*=\pm1$ 
and the Legendre function $P^{\cal M}_{\cal J}(\pm 1)=0$ for ${\cal M}\ne
0$. 
Also they do not contribute when the azimuthal angles of the polarization
direction $\phi^*$ and the emission direction $\phi$ are the same
($\Delta\phi=0$). As noted above, the total number of angular 
reduced response 
functions for nucleon knockout from the $d_{3/2}$ shell of $^{39}$K is 72, 
which is too large for a complete presentation and therefore in the
following 
we only show figures for the most relevant. 
The complete set of figures is presently available from the 
authors~\cite{www97}.

The polarized responses
 $W^{K(\pm)}_{\cal JM}$ 
and  $\widetilde{W}^{K(\pm)}_{\cal JM}$ 
depend on the three variables $(q,\omega,p)$, since the missing energy is 
determined. In order to study their dependence on all of the variables, we 
have computed the entire set of responses as functions of the missing
momentum 
$p$ for nine values of $(q,\omega)$. In each one of the following figures
we 
show a particular response function and each figure is composed of nine
panels
corresponding to the values of $(q,\omega)$ given in Table~1. 
In each one of the rows in the table the value of $q$ is fixed to one
of the three values  $q=300$ MeV/c
(regions Ia, Ib, Ic), $q=500$ MeV/c (regions IIa, IIb, IIc) 
or $q=700$ MeV/c (regions IIIa, IIIb, IIIc). We show
three values of $\omega$ for each $q$. The second column corresponds 
approximately to the quasielastic peak (QEP) defined by $p'\simeq q$. 
The first column corresponds to slightly below the peak, $p'<q$,
while the third column is in the region above the QEP, $p'>q$. 
The values of the momentum $p'$ and kinetic energy $\epsilon'$ 
of the ejectile for each of the panels are also shown in Table~1.

\begin{table}
\begin{center}
\begin{tabular}{|l|l|l|} \hline
{\bf Ia}          & {\bf Ib}          & {\bf Ic}        \\
$q=300$ MeV/c           & $q=300$ MeV/c           & $q=300$ MeV/c       \\
$\omega=40$ MeV       & $\omega=56$ MeV       & $\omega=80$ MeV         \\
$p'=244.3$ MeV/c        & $p'=301.6$ MeV/c        & $p'=372.6$ MeV/c    \\
$\epsilon'=31.3$ MeV  & $\epsilon'=47.3$ MeV  & $\epsilon'=71.3$ MeV \\
\hline
{\bf IIa}         & {\bf IIb}         & {\bf IIc}       \\
$q=500$ MeV/c           & $q=500$ MeV/c           & $q=500$ MeV/c       \\
$\omega=110$ MeV      & $\omega=133.5$ MeV    & $\omega=160$ MeV        \\
$p'=447.6$ MeV/c        & $p'=499.7$ MeV/c        & $p'=553.9$ MeV/c    \\
$\epsilon'=101.3$ MeV & $\epsilon'=124.8$ MeV & $\epsilon'=151.3$ MeV \\
\hline
{\bf IIIa}        & {\bf IIIb}        & {\bf IIIc}      \\
$q=700$ MeV/c           & $q=700$ MeV/c           & $q=700$ MeV/c       \\
$\omega=180$ MeV      & $\omega=241$ MeV      & $\omega=300$ MeV        \\
$p'=592.3$ MeV/c        & $p'=699.9$ MeV/c        & $p'=794.6$ MeV/c    \\
$\epsilon'=171.3$ MeV & $\epsilon'=232.3$ MeV & $\epsilon'=291.2$ MeV \\
\hline
\end{tabular}
\end{center}
\caption{\protect\em Values of 
the momentum and energy transfer $(q,\omega)$, 
and of the momentum $p'$
and kinetic energy $\epsilon'$ 
for which the reduced response functions
have been computed.}
\end{table}

First notice that the Schwandt optical potential we use was originally 
parametrized for elastic scattering of protons with energies 
between 80 and 180 MeV, and therefore it has been extrapolated 
in some of the kinematic regions. Also the daughter nucleus
$^{38}$Ar was not used in the analysis of Schwandt {\em et al.}, but it 
is in the mass range of the nuclei originally used. One of the aims of
this work is to analyze the sensitivity of the different responses
to the FSI model, and, in particular, in each one of the 
panels in Figs.~\ref{L00}--\ref{TL'30}
we show four curves, each one corresponding to 
a particular model for the final proton state. 
Specifically, the solid lines have been computed 
with the complete optical potential, while for the dashed lines
we have set the spin-orbit potential (SOP) equal to zero
(both real and imaginary parts). Thus comparison between solid 
and dashed lines shows the effect of the SOP in the responses. 
These two curves do  not include a Perey factor in the wave function,
and therefore we also show with dot-dashed lines the 
same responses computed with the total potential and with 
Perey parameter $\beta=0.85$ in the wave function. Finally, 
the dotted curves correspond to the PWIA, that is, they
do not include FSI. 

In order to get a feeling about the
relative order of magnitude of the 72
responses it is convenient to choose some 
quantity which characterizes the importance 
of each response. 
The total variation of a response 
defined as the maximum minus the minimum values 
$\Delta W^{K(\pm)}_{\cal JM} \equiv
W^{K(\pm)}_{\cal JM}({\rm max}) -
W^{K(\pm)}_{\cal JM}({\rm min}) $,
that is always positive, is a quantity that is very useful for
that purpose. Specifically, we take the variation of the unpolarized
longitudinal $\Delta W^{L(+)}_{00}$ as reference 
and therefore give the others as a percentage 
of this response, so the variation 
of the unpolarized longitudinal  response is 100\% by definition.
We have computed the total variation of all of the responses
using the complete optical potential 
for the nine kinematic conditions and the results are shown 
 in Tables 2--4 for $q=300$, $500$ and $700$ MeV/c, respectively. 
Note that under the columns for the $L$, $T$ and $T'$ responses 
both kinds of responses (with and without tilde) are displayed
because in these cases there is no confusion: the
$L$ and $T$ responses without (with) tilda  are always of
even (odd) rank, while the opposite happens for the $T'$ 
responses. 

With reference to the PWIA responses, 
only the 32 responses given in Eqs. (\ref{169}--\ref{174}) survive
in that limit.
Therefore, a total of 40 of the 72 responses are exclusively due to the 
FSI, and then it is expected that they will be especially sensitive
to the details of the interaction, as we shall  show below for some of the
more important cases.

%----------------------------------------------------
\subsection{Detailed discussion of selected responses}
%-----------------------------------------------------

%---------------------------
\subsubsection{L-responses}
%---------------------------

\paragraph{Unpolarized response.} 

The unpolarized 
response $W^{L(+)}_{00}$ displayed in Fig.~\ref{L00} serves to set
the scale of the remaining responses. The three panels in 
the middle row of Table~1 at $q=500$ MeV/c are representative 
and well within the range of validity of the Schwandt potential. The one 
in the middle at $\omega=133.5$ MeV (region IIb) approximately corresponds
to the QEP. The FSI (solid lines) produce a reduction of the 
plane-wave results (dotted lines) 
of about 35\%. When we set the SOP to zero, the response increases by a
few percent (dashed lines); also we see that the effect is bigger 
for higher energy. The introduction  
of a Perey factor slightly reduces the total response (dot-dashed lines),
but the effect is very small. A significant effect of the FSI
with respect to the PWIA is seen in the extreme cases, 
$q=300$ MeV/c, $\omega=40$ MeV (region Ia) and 
$q=700$ MeV/c, $\omega=300$ MeV(region IIIc), although 
these results should
be viewed with caution, being far outside the range of validity 
of the potential. Indeed, in the former case the real attractive part of
the potential is too large, while in the latter the imaginary 
absorptive part is too large. In region IIIa 
($q=700$ MeV/c, $\omega=180$ MeV) the spin-orbit interaction 
produces an increment of the response (the opposite 
to what happens in the other cases). The effect of the Perey factor 
decreases with the energy because it only depends on the real part 
of the potential, which also decreases with the energy
(this conclusion is also applicable to the remaining responses). 

\paragraph{Vector response.} 

The vector longitudinal response $\widetilde{W}^{L(-)}_{11}$, 
shown in Fig.~\ref{L11tilde}, is zero in PWIA.
Thus it is due only to the FSI and expected to be more sensitive to details
of the potential. In addition, it is not as large as the unpolarized one,
but 
we find that it is large enough to make its extraction feasible --- in
fact, 
in region IIb at the maximum $p\simeq 150$~MeV/c, its magnitude is about
24\% of $W^{L(+)}_{00}$ 
(see Table~3). In regions Ib,c and IIIb,c, it is even bigger. 
For instance, we see in Tables~2 and 4 that the variation 
$\Delta W^{L(-)}_{11}$ in regions Ib and IIIb is about 56\%
and 44\%, respectively, of the unpolarized longitudinal.

Without the spin-orbit interaction this response is much 
larger and in some cases its behaviour as a function of $p$ is even
different. 
Specifically, in the QEP region (second
column of Fig.~\ref{L11tilde}) 
the SOP produces a small shift to the right, which is
more important for low $q$. The same happens in the region to the right 
of the QEP. However, more important effects are seen below 
the QEP, where the response is significantly reduced by the 
spin-orbit interaction and its shape is smoothed and broadened. 
The effect is more pronounced at high energy; for instance, at $\omega=180$
MeV
(region IIIa) the SOP produces a reduction of the peak of more than
60\%. 

This response contributes the most to the cross section when the 
polarization is normal to $q$, namely at $\theta^*=90^\circ$ because
it is multiplied
by a Legendre function $P^1_1(\cos\theta^*)=\sin\theta^*$, and for 
$\Delta\phi=\pm 90^0$. Therefore we expect the spin-orbit interaction 
to be important under these conditions. 
For example, for coplanar kinematics ($\phi=0$) and normal (N) polarized
$^{39}$K ($\theta^*=\phi^*=90^\circ$), where 
$\Delta\phi=\phi-\phi^*=-90^\circ$.

\paragraph{Quadrupole responses.}

In this case the PWIA responses $W^{L(+)}_{2\cal M}$ are nonzero 
and in general they are large, as seen in Tables 2--4.
The biggest one corresponds to ${\cal M}=0$, 
shown in Fig.~\ref{L20},
which contributes significantly 
for $L$-polarization: for instance, its size is about one-half of the 
unpolarized response in region IIb
(see Table~3). For low $p$ this response is in general 
negative and here the PWIA provides insight into why this happens. 
  From Eq.~(\ref{169}), we see that this response is 
proportional to the angular function $h^0_{2200}(\theta)$
and to the radial function  $m^S_2(p)$.
From its definition, eqs. (\ref{76},\ref{77})
it is easy to  see that the function $h^0_{2200}(\theta)$
is proportional to $P^0_2(\cos\theta)=\frac12(3\cos^2\theta-1)$,
while the radial function $m^S_2(p)$,
defined by Eq.~(\ref{147}),  is proportional to 
the factor $A_2$, given in ref. \cite{Ama97}, which is proportional to 
a negative 3-$j$ coefficient. Therefore, this response 
includes a factor $1-3\cos^2\theta$. Below the QEP
we have $p'<q$ and therefore, for $p$ small, $\np$ is almost
anti-parallel to $\nq$ because $\np+\nq=\np'$, and so the angle 
$\theta$ is near $180^\circ$ and therefore $1-3\cos^2\theta\simeq-2$
is negative and the total response is negative. 
As $p$ increases, the angle $\theta$ quickly reaches a value near 
$90^\circ$ degrees, and so the factor $1-3\cos^2\theta\simeq 1$ changes 
its sign. This change in sign is almost immediate for QE conditions
because $p'\sim q$, while above the QEP $p'>q$ and for $p$ small 
now $\theta\sim 0^\circ$ and again the response is negative
and changes its sign as $p$ increases. 
An interesting feature of this response is that 
the effect of the FSI is about the same as that in the unpolarized case
in the QEP and above it, but below the QEP 
the FSI produce an increase of the response that is significant at
low $q$. 
In fact, for fixed $q$ (i.e., for each row in Fig.~\ref{L20}), 
we see that there is an inversion of 
the PWIA and DWIA responses at some point below the QEP
between regions a and b (first and second columns of Fig.~\ref{L20}).
At $q=500$ MeV/c and QE conditions (region IIb) the
effect of the SOP is somewhat bigger than in the unpolarized case.

\subsubsection{T responses}
%-----------------------------------

\paragraph{Unpolarized response.} 

The unpolarized response $W^{T(+)}_{00}$ 
shown in Fig.~\ref{T00} presents
features that are similar to those 
seen in the longitudinal one, with the exception that it is rather
insensitive
to the spin-orbit part of the interaction. 
As we can see in the figure, the solid and dashed lines are very 
similar. An interesting characteristic can be seen if we compare 
the behaviour of the variations 
$\Delta W^{T(+)}_{00}$ both in DWIA (Tables 2--4) and PWIA. 
In DWIA the total variation --- relative 
to the longitudinal --- increases with energy for $q=500$ MeV/c
from $\sim 100\%$ (region IIa) to $\sim 104\%$ (region IIc), see table 3,
while in PWIA we have verified that 
the reverse occurs: there is a decrease from 
$\sim 101\%$ to $\sim 91\%$. 
Therefore one expects an increase of the $T/L$ ratio with
$\omega$ due to FSI, whereas in pure PWIA the $T/L$ ratio
decreases with energy. This means that the FSI is stronger in the 
L-response than in the T-response. As we can see comparing Figs.~1 and 4,
this is due to the fact that the SOP produces an additional reduction 
of the L-response, while the T-response appears to be insensitive to the 
SOP.

\paragraph{Vector response}
The same behaviour of the small SOP effects in the unpolarized transverse
 response is also observed
in the polarized $T$ responses. For instance, the very large
effect observed in the 
$\widetilde{W}^{L(-)}_{11}$
response does not happen for $\widetilde{W}^{T(-)}_{11}$
(not shown). 
We conclude that the $L$ responses are sensitive to the SOP,
while the $T$ responses are almost independent of it. 
This may be due to the different spin-dependence of the 
corresponding electromagnetic operator. In the $L$ response the
electromagnetic operator in leading order involves the spin-independent 
charge operator,
which cannot flip the spin. Therefore the transition matrix
element of the $L$ response  is sensitive to the change of the
spin part of the wave function induced by the SOP. 
On the other hand, the transverse current is proportional to 
the Pauli spin matrices in leading order,
and therefore it is less dependent on the relative change of the
$l+\frac12$ and $l-\frac12$ components of the partial waves. 
With regard to size, the $\widetilde{W}^{T(-)}_{11}$ response is about
25\% of the unpolarized longitudinal for $q=500$ MeV/c and
is much bigger for $q=700$ MeV/c (see Tables 2--4).
We have observed that the vector transverse response approximately scales 
with the vector longitudinal one when this is computed without
the SOP and that the scaling factor is about the same as in the 
unpolarized case (and this is the reason for not showing 
figures for this response).

\subsubsection{TL responses}
%------------------------------------------------

\paragraph{Unpolarized response.}

The $W^{TL(+)}_{00}$ shown in Fig.~\ref{TL00}
is an interference between $L$ and $T$ multipoles
and thus some similar behaviour to the responses discussed above is
expected,
although the interference can produce a reduction or enhancement of 
some of the features. In fact, in PWIA the $TL$ response
is of first-order in $\delta=\eta\sin\theta$, as we can see from
Eq.~(\ref{171}), and it is due to the interference between 
charge and convection, and spin-orbit and magnetization terms. 
In the QEP, the $TL$ strength is around 50\% of the longitudinal
(Tables 2--4).
For $q=300$ MeV/c the $TL$ response is bigger than the 
$T$ response. This can be understood by looking at the expressions for the 
responses, Eqs.~(78---83) of reference \cite{Ama97}: in leading order
the $T$ response is proportional to the single-nucleon response 
$w^T\sim 2J_m^2=2\tau G_M^2$, while the 
TL response is proportional to 
\[
w^{TL}=2\sqrt{2}(\rho_cJ_c+\rho_{so}J_m)\delta=
2\sqrt{2}\left(G_E^2+\frac{2G_M-G_E}{\sqrt{1+\tau}}
\frac{\kappa}{2}\sqrt{\tau}G_M\right)\delta.
\]
For $q=300$ MeV/c we have $\tau\sim \kappa^2$, while for the maximum of
the momentum distribution at $p\sim 150$ MeV/c, $\theta\sim 90^\circ$,
we have $\delta\sim\kappa$ and therefore 
\[ \frac{w^T}{w^{TL}}\sim \frac{2\kappa\mu_P^2}{\sqrt{2}}\sim 0.9.
\]
Thus, for low $q$ the $TL$ response is bigger than the $T$ response,
although the latter is of zeroth-order in $\delta$, while the $TL$ one 
is of first-order in the expansion.
Of course, the currents contain terms which
are $(q,\omega)$-dependent, whose magnitude has to
be taken into account. For high $q$ the factor $\tau$ is
bigger and the $T$ response dominates over the $TL$ one. 

With respect to the FSI effect, it is bigger than the one found 
in the $T$ and $L$ responses. 
The reduction in all cases is more than 50\%. 
This suggest that the convection and spin-orbit electromagnetic matrix
elements are more sensitive to the dynamics of the FSI. Also 
this response is more sensitive to the SOP than the longitudinal one.
In fact, if the SOP is set to zero, the FSI effects are almost the same 
as in the $L$ and $T$ responses. Therefore it appears to be the 
{\em spin-orbit 
potential} that is responsible for the very significant reduction of this 
response when compared with the other two. 

In order to understand this feature better, we have performed a calculation
of the responses with the SOP set to zero and without spin-orbit
and convection pieces in the current. As a result, we have checked
numerically that the interference between charge and magnetization
goes to zero when the SOP does so. This happens for all of the 
polarized $TL$ responses. In addition, when the 
SOP is turned on, the zeroth-order contribution is nonzero
and negative for $q>500$ MeV/c at the QEP, providing the reason 
why the FSI are so large in this response. 

\paragraph{Vector responses.}

There is no counterpart of the $W^{TL(-)}_{10}$ response in the $T$ 
and $L$ cases. 
We show this response in Fig.~\ref{TL10}.
As seen in Eqs.~(\ref{66b},\ref{67b}) the vector responses without a tilde 
contribute only to the $\widetilde{W}^{TL}$ structure function,
which is nonzero only for out-of-plane emission (see Eq.~(\ref{RTL})). 
This vector response presents an oscillatory behaviour, its magnitude 
being  $\sim 26\%$ of the unpolarized longitudinal in region IIb.
From the figure it is evident that  this response is very sensitive  
to the interaction, the results with and without SOP
being completely different in all of the kinematical regions here studied.
In fact, the depencence of this response on the SOP is much more
dramatic than for the case of the vector longitudinal
(compare Figs. \ref{L11tilde} and \ref{TL10}), and therefore such
 big effects, even changing the sign in many cases,
could in principle be measured if a separation of this response 
could be performed.

\paragraph{Quadrupole responses.}

The responses 
$W^{TL(+)}_{2\cal M}$ are 
nonzero in PWIA and, under that approximation, the
relation between 
$W^{TL(+)}_{2\cal M}$ 
and the corresponding 
$W^{L(+)}_{2\cal M}$ 
and 
$W^{T(+)}_{2\cal M}$ 
is the same as in the unpolarized case. 
The FSI effects in the 
$W^{TL(+)}_{20}$ response shown in Fig~\ref{TL20}
are stronger than in the case of 
$W^L_{20}$
and $W^L_{20}$,
as can be seen comparing Figs. 3 and 7. This is again a consequence
of the different sensitivities of these responses to the 
spin-orbit interaction (compare the dashed curves
in Figs. 3 and 7). The reason for this behaviour is again that the SOP
generates a contribution from zero-order terms 
(interference between charge and magnetization) in the current 
matrix elements, which would be zero if there were no SOP. 

The FSI effects 
in the $W^{TL(+)}_{20}$ 
response also
 are very dependent on the choice of kinematics. 
For $q=500$ MeV/c at the QEP (region IIb),
the FSI reduce the response by about 60\%
and below the QEP in region IIa, they produce a change of sign.
The SOP importance is very large at low $q$,
but small above the QEP in region IIc. 
On the other hand, at the QEP the importance of FSI 
and SOP strongly increases  with $q$: at $q=300$ MeV/c the
total effect is about 10\%, while the response is drastically 
reduced for $q=700$ MeV/c.

\subsubsection{T$'$ responses}

Polarized electrons are required for the primed responses
$T'$, $TL'$. There are a total of 24 reduced response functions of this
kind which we now discuss.
Note that there is no $T'$ response in absence of 
nuclear polarization. In fact, for ${\cal J}={\rm even}$ the $T'$ 
responses are only 
of the tilde type, and therefore all of them are zero for ${\cal M}=0$.

\paragraph{Vector responses.}

The  $W^{T'(-)}_{10}$ response shown in Fig.~\ref{T'10}
 is nonzero in PWIA and 
has no counterpart in the set of unprimed responses, since there is no
vector $T$ response of this kind. From Eq.~(\ref{173}) 
we see that in leading order in $\delta$ this response 
is expected to be of the same order of magnitude as the $T$
responses and in fact in PWIA this response is comparable 
with the $T$ unpolarized one,
and the same happens in DWIA, as we can see by inspection of
Tables 2--4. The FSI produce a reduction 
that in general is smaller than that found for
$W^{T(+)}_{00}$; however we find several exceptions
to this statement, the most important one being that in region IIIa
the FSI only slightly affect this response. In this region the only
noticeable 
FSI effect is found for parallel emission ($\theta'\sim 0$), although
for higher missing momentum ($p'\sim 200$ MeV/c) the response
is almost independent of the interaction. 
In addition for $q \ge 500$ MeV the SOP effect is negligible. 
Therefore
this could be a good place to look for other model dependences
such as on the specifics of the single-nucleon current or nuclear
structure.
A second exception is in region Ia, where the FSI enhance
and produce a shift of the response, although we are far away
from the range of applicability of the potential.

With respect to the 
$W^{T'(-)}_{11}$ 
response (Fig.~\ref{T'11}),
 one would expect that it should in general be bigger than 
$\widetilde{W}^{T(-)}_{11}$
which is zero in PWIA, but we find that in many cases the 
FSI produce a significant reduction of the PWIA results for this response,
the effect being more than 60\% in the QEP region,
where the variation $\Delta W^{T'(-)}_{11}$ of this response is always 
smaller than the corresponding $\Delta W^{T(-)}_{11}$ (see Tables 2--4).
On the other hand, in regions IIa,c, this variation is $\sim 50\%$ 
of the unpolarized longitudinal case, and it is above
100\% in regions IIIa,c. The reason why this 
response is so small at the QEP can be seen by 
looking at its  behaviour for fixed $q$, as a function of $\omega$
(Fig.~\ref{T'11}).  There is a tendency for this response to change 
from positive at lower $\omega$ to negative at some point slightly 
above the QEP. Comparing the DWIA and PWIA results we also see a 
general tendency for the former to be reduced in magnitude with respect 
to the latter, where this reduction is seen to be significant below the 
QEP and is larger for low
$q$. On the other hand, above the QEP (regions Ic, IIc) the
mean (negative) peak of this response is almost not affected by
FSI, so these could be good places to investigate other aspects 
of the reaction. This trend does not apply for $q=700$ MeV/c, $\omega=300$
MeV
(region IIIc), but here again we are far away from the validity of the
potential parametrization. As in the case of the T response, this response
is insensitive to the SOP.

\paragraph{Octupole responses.}

The  $W^{T'(-)}_{30}$ response shown in Fig.~\ref{T'30}
is comparable to the $W^{T'(-)}_{10}$
one in both PWIA and DWIA, and over the entire kinematical range
considered.
In some places there are drastic changes due to the FSI with respect to the
PWIA. For example, in region IIa below the QEP
this response is almost opposite to the 
$W^{T'(-)}_{10}$
in PWIA, that is reduced by FSI by less than 20\%, while the 
$W^{T'(-)}_{30}$ 
is almost canceled by FSI. On the other hand, 
in region IIb, the FSI effect provides a reduction of
around 20\%. If we go further in energy (to region IIc), where
the PWIA presents an oscillatory behaviour, the FSI yield a reduction 
by 50\% at the maximum, and almost do not affect the first minimum. 
We observe a similar behaviour in region Ic where the reduction 
due to FSI is even bigger, around 75\%.
Again, the SOP effect is negligible in this response.

The $W^{T'(-)}_{31}$ response 
is shown in Fig.~\ref{T'31} and is also important.
It reaches its greatest values above the QEP
(see also Tables 2--4). For instance, 
its variation is about 25\%,  
50\% and 120\% of the unpolarized longitudinal 
in regions Ic, IIc, and IIIc, respectively.
At the QEP, its magnitude is around 70\% of $W^{T'(-)}_{11}$,
and the FSI effects are almost identical in both
of them, in general very strong for low $q$ (producing
even a change of sign) and moderate at $q=700$ MeV/c
(with a reduction of $\sim 65\% $).
Below the QEP in region IIa this response is around 80\%
of the 
$W^{T'(-)}_{11}$
and about one half of 
$W^{T'(-)}_{10}$,
the FSI effects being smaller than in the 
$W^{T'(-)}_{11}$
response. FSI effects are also seen to be very different at the QEP 
upon comparing the ${\cal M}=0$ and ${\cal M}=1$ odd $T'$ responses. 
It is important to note that the identification of places where FSI 
effects are very different, as in the case of the 
$W^{T'(-)}_{11}$ and
$W^{T'(-)}_{31}$
in region IIa, could play an important role in this type of study where 
the goal is to isolate such effects.
The remaining octupole responses
$W^{T'(-)}_{32}$ and 
$W^{T'(-)}_{33}$
(not shown in the figures)
tend to be smaller than the one discussed above, although the former
could be noticeable in some cases, while the latter is negligible. 

In all cases we find  that the  influence
of the SOP in the $T'$ responses is small, as in the $T$ case,
thus making it possible to use these transverse responses to study 
the central part of the interaction and to use
the responses which are sensitive to the SOP,
such as the $L$ or $TL$ cases, to study the spin-dependent part
of the interaction. 

\subsubsection{TL$'$ responses}

\paragraph{Unpolarized response.}

The ``fifth response function'' 
$W^{TL'(+)}_{00}$ 
(shown in Fig.~\ref{TL'00})
is zero in PWIA. It can be separated by performing an asymmetry 
measurement  by flipping the electron helicity in experiments with
unpolarized nuclei, therefore providing an alternative way of
studying the FSI. First we note that its size is large and comparable with
the 
unpolarized $TL$ response, with which it is related by 
Eqs.~(\ref{67}--\ref{70}). In fact, the $TL$ and $TL'$
responses differ in that they select different combinations
of the real or imaginary parts of the interference multipoles 
in Eqs.~(\ref{46},\ref{47}). Specifically, the 
$W^{TL(+)}_{00}$ 
and  $W^{TL'(+)}_{00}$ contributions 
contain the real and imaginary parts, respectively, of the same
combination
of multipoles (see Eqs.~(\ref{58},\ref{59})). Therefore their
magnitudes are roughly determined by the averages
$\langle\cos\Delta\delta\rangle$
and
$\langle\sin\Delta\delta\rangle$
of 
$\cos[{\rm Re}(\delta_{l'j'}-\delta_{lj})]$
and 
$\sin[{\rm Re}(\delta_{l'j'}-\delta_{lj})]$, respectively, 
that is, 
of the differences between the real phase-shifts for different
partial waves. The magnitudes of these averages are determined
by factors involving the weighting amplitudes 
$|C_{\sigma'}E_{\sigma}|$
and $|C_{\sigma'}M_{\sigma}|$
in Eqs.~(\ref{67}--\ref{70}). The intercomparison 
between $TL$ and $TL'$ responses can  therefore provide 
information about the mean real phase-shift deviation. 
Of course, within the average the relation
$\langle\cos\Delta\delta\rangle^2+
 \langle\sin\Delta\delta\rangle^2=1$
is in general not true, although for QE conditions it is 
approximately satisfied.
In fact, the values of the pair
$(W^{TL(+)}_{00},W^{TL'(+)}_{00})$
at the peak in regions Ib, IIb and IIIb
are $\sim(70,40)$, (70,45) and (30,70) [GeV$^{-1}\times 10^{-3}$]
and therefore the combination 
$\left[W^{TL(+)}_{00}\right]^2+\left[W^{TL'(+)}_{00}\right]^2
\simeq{\rm constant}$.
Outside the QEP this relation does not hold, but in any case 
the two responses are always comparable in magnitude. 
Moreover, the effect of the SOP is of similar importance, 
$\sim 20\%$ in both responses. 

With regard to their behaviour as functions of the kinematic variables, the
$TL$ and $TL'$ responses 
have different structures, the most significant variations
being found below the QEP and at low energy. 
It is interesting to note that the behaviour of the 
$W^{TL'(+)}_{00}$ response
is more similar to that of the 
$\widetilde{W}^{L(-)}_{11}$
and $\widetilde{W}^{T(-)}_{11}$
responses. Our results show that the effect of the SOP
in the 
$W^{TL'(+)}_{00}$ 
and $\widetilde{W}^{L(-)}_{11}$
responses is similar, and it presents a tendency 
to modify the response in the same way. 
It is also noteworthy that when the SOP is set to zero, the
three responses 
$W^{TL'(+)}_{00}$,
$\widetilde{W}^{L(-)}_{11}$
and $\widetilde{W}^{T(-)}_{11}$
acquire the same shape and that 
$W^{TL(-)}_{10}$ also shows the same tendency 
(compare Figs.~\ref{L11tilde}, \ref{TL10} and
\ref{TL'00}). This is related
to the fact that all of these responses are given by imaginary parts
of the single transition responses for each partial wave. 

\paragraph{Vector responses}

The $TL'$ vector responses are large in magnitude and in all cases
considered they are comparable to the $W^{L(+)}_{00}$ response. 
  From Tables 2--4 we see that their importance increases with the
momentum transfer, and for high $q$ the $\widetilde{W}^{TL'}_{11}$ 
response is the dominant one, being around 160\% and 
220\% of the unpolarized longitudinal case for $q=500$ and 700 MeV/c,
respectively.
The $W^{TL'(-)}_{10}$ response
is shown in Fig.~\ref{TL'10}, and it
presents a behaviour which is opposite in sign to 
${W}^{T'(-)}_{11}$, both with and without FSI, but the
effect of the SOP is different at the QEP (regions Ib, IIb, IIIb).
For instance, in region IIb the introduction of the SOP changes
the sign and peak position, and in region IIIb the response is
negligible without SOP, while its inclusion produces an
increase of the absolute strength of the PWIA results
by more than  100\%.
On the contrary, in region Ic the FSI effects are negligible. 
The  $W^{TL'(-)}_{11}$ and
$\widetilde{W}^{TL'(-)}_{10}$ responses
 are very large in magnitude and almost 
opposite in sign.
It is interesting that the  
$\widetilde{W}^{TL'(-)}_{11}$ response,
not shown in the figures, 
has a behaviour that is very similar to 
$W^{L(+)}_{00}$,
although opposite in sign, and that the effect of the FSI and SOP
 is also similar in  these two responses. This is related to the 
fact that all of these responses are composed of real parts
of responses for single multipole
transitions 
$R^L_{\sigma'\sigma}$,
$R^{T1}_{\sigma'\sigma}$,
$R^{T2}_{\sigma'\sigma}$,
$R^{TL1}_{\sigma'\sigma}$, and
$R^{TL1}_{\sigma'\sigma}$
in Eqs.~(\ref{61},\ref{63},\ref{67},\ref{69}), and that all of 
them are dominated by the zeroth-order static component of the 
electromagnetic current.

\paragraph{Quadrupole responses.}

The 
$W^{TL'(+)}_{20}$, shown in Fig.~\ref{TL'20},
has no $T'$ counterpart. This response is bigger 
than the fifth response function below the QEP, but is smaller at
 and near the QEP. 
This response shows a large sensitivity to the  SOP,
which completely changes its behaviour in many cases, 
especially at the QEP.
Specifically, the SOP produces the tendency to reduce this 
response to negative values for quasielastic conditions. In general
all of the 
$W^{TL'(+)}_{2\cal M}$ 
and $\widetilde{W}^{TL'(+)}_{2\cal M}$ responses
have almost the same magnitude in region IIb, and they are
about twice that of the 
transverse 
$\widetilde{W}^{T'(+)}_{2\cal M}$
(see Tables 2--4),
which could help in making 
an experimental separation of these quadrupole responses feasible.
In general all of the quadrupole $TL'$ responses are very
sensitive to the SOP, in contrast to what happens for 
the quadrupole $T'$ responses. 

\paragraph{Octupole responses.}

There are four
$W^{TL'(-)}_{3\cal M}$ 
and three
$\widetilde{W}^{TL'(-)}_{3\cal M}$ 
octupole responses and in general all of them are big compared
with the fifth response function,
the biggest ones being for ${\cal M}=0,1$, as seen by inspection
of Tables 2--4. In fact, 
$W^{TL'(-)}_{30}$, shown in Fig.~\ref{TL'30}, is the response with 
the biggest variation in regions IIa,c and IIIa,c, and 
$W^{T'(-)}_{30}$ is the second response of importance in this ranking.
The
$W^{TL'(-)}_{30}$, 
$W^{TL'(-)}_{31}$ 
and $\widetilde{W}^{TL'(-)}_{32}$ responses
are strongly affected by FSI, both in size and sign at QE conditions,
and the last two are significantly suppressed by the FSI. The SOP
does not play a particularly important role in these responses, with 
some exceptions. In regions IIa, IIc and IIIa, the 
$W^{TL'(-)}_{30}$ response is larger in magnitude than the unpolarized 
longitudinal response
even with the inclusion of FSI and it is rather insensitive 
to the SOP. Also $W^{TL'(-)}_{31}$ 
is large in these regions, where it is similar in shape 
and magnitude to 
$W^{T'(-)}_{30}$. In summary, the large magnitude of these octupole
responses
 indicates that they could possibly be separated through a sequence 
of asymmetry measurements.

%------------------------------------
\subsection{Relativistic corrections}
%------------------------------------
 
In this section we briefly discuss the importance of the different terms
in the first-order expansion of the relativistic current
in powers of $\delta=\eta\sin\theta$ \cite{Ama97}.
In particular, the goal is to study the importance of the first-order 
terms which we call spin-orbit (in the longitudinal)
and convection (in the transverse), but which differ
from the terms usually used in non-relativistic calculations
in that they maintain the full dependence on $(q,\omega)$
in the relativistic current 
(see the definitions in Eqs.~(\ref{157}--\ref{160})), and therefore
they are more suited for use at high momentum transfer 
near the QEP. Also the spin-orbit charge (SOC) term is usually
not included in non-relativistic calculations, but as seen 
in Eq.~(\ref{171}), it contributes to the single-nucleon
$TL$ response at the same level as the convection current and therefore
its inclusion in the longitudinal component
can be important for properly describing the $TL$ responses
at high momentum transfer, as was found in Ref.~\cite{Ama97}
in the inclusive polarized case. Moreover, the FSI can affect each one of 
these contributions in a different way, as may be reflected especially in
the responses which are zero in PWIA.

In the following we show only two figures to illustrate the most interesting
cases.
As in the last section, we show a separate response function in the nine
panels
corresponding to the same choices of kinematics made above.
Now the solid line 
in Figs.~\ref{soTL00}--\ref{soTL11tilde}
represents the total response computed in DWIA with 
the complete optical potential. With dashed lines we show  
the responses computed in leading order in the expansion, and therefore
the dashed lines only include charge plus magnetization 
(zeroth-order) terms. Finally, the dot-dashed lines include, in addition,
 the convection current, and thus they are representative of what can
be found in a standard non-relativistic calculation (with the important
difference that we include the relativistic $(q,\omega)$
behaviour which is significant at high $q$). In other words, the effect
of the spin-orbit term in the longitudinal current is to transform 
the dot-dashed curves to the solid ones. 

We now discuss the importance of the different parts
of the electromagnetic current in the 
case of the  $TL$ responses.
In PWIA these responses are entirely due to interferences between 
charge and convection, and spin-orbit and magnetization terms, 
and as a consequence the first-order terms of the current
are crucial when describing these responses. Of course,
when FSI are turned on,
the cancellation of the zeroth-order terms is not exact
and the interference between charge and magnetization is nonzero 
with FSI (dashed lines in the figures),
although its magnitude is comparable with the total response
or smaller in some cases.
As stated above, we have checked that the zero-order
contribution to these responses goes away when the SOP
is set to zero, and this is the reason why the $TL$ responses are so
sensitive
 to the spin-orbit part of the interaction.

The  $W^{TL(+)}_{00}$, 
shown in Fig.~\ref{soTL00},
is especially sensitive to the 
first-order terms.
The zeroth-order terms (interference between charge and magnetization
pieces)
are shown with dashed lines and they are zero in the absence of 
a spin-orbit interaction in the optical potential. 
The importance of the zeroth-order terms increases with the momentum
transfer in all kinematic regions. For $q=300$ MeV/c this
contribution is almost negligible, while for $q=700$ MeV/c
it is of the same order of magnitude as the total responses and opposite 
in sign. The increasing importance of the zeroth-order term is closely
related with what we found in last section --- the effect of
the SOP increases with $q$ in this response and the inclusion 
of the SOP potential in the FSI prevents the spin cancellations
from occurring in the  zeroth-order term.
Shown with dot-dashed lines is the response computed including only 
charge, magnetization and convection terms
in the electromagnetic current. These are the expected results
if a standard non-relativistic treatment is used to compute
the responses (no spin-orbit correction to the longitudinal piece).
The effect of the convection term is to completely change the behaviour 
of this response with respect with the zeroth-order, 
even changing the sign for $q \ge 500$ MeV/c. 
When we include also the SOC piece, we obtain the solid lines in
the figure, thereby showing the importance that this correction has 
in the $TL$ response. 
The SOC contribution also increases
with $q$. The effect of this term is to increase the total 
(dot-dashed)
response by $\sim 15\%$, $\sim 75\%$ and $\sim 200\%$ 
at the peak for $q=300$, 500 and 700 MeV/c, respectively, and QE
conditions (regions Ib, IIb, IIIb).
The effect is even  bigger below the QEP, in regions Ia,b,c.
As seen in Eqs.~(\ref{158},\ref{160}),
the spin-orbit charge is proportional to 
$\kappa/\sqrt{1+\tau}$, which increases with $q$,
while the convection piece is proportional to 
$\sqrt{\tau}/\kappa$, which slightly decreases with 
$q$. Therefore the relative importance of the SOC versus
the convection current increases with $q$, as seen in the 
plots. 
Accordingly we conclude that any non-relativistic model 
which does not include the spin-orbit term
significantly underestimates this response at a typical
value of $q=500$ MeV/c and QE conditions.

The zeroth-order term is important in the 
$W^{TL(-)}_{1\cal M}$ responses. For QE conditions, 
the convection effect is crucial at $q=300$ MeV/c, and decreases with 
$q$. In region IIIb spin-orbit
and convection effects are only around 20\% of the total response
in 
$W^{TL(-)}_{10}$
and smaller in $W^{TL(-)}_{11}$. 
These effects are much more important below the QEP, although
in this region these responses are small compared with the 
$W_{00}^{L(+)}$. On the other hand, the first-order terms are more 
important in the $\widetilde{W}^{TL(-)}_{11}$ case shown in 
Fig.~\ref{soTL11tilde}, which is exactly zero in PWIA.
The zeroth-order terms are small in 
$\widetilde{W}^{TL(-)}_{11}$
which presents the same trends as the unpolarized response,
$W^{TL(+)}_{00}$, and both responses are of the same order for
$q=700$ MeV/c (see Table~4). For instance, in region IIIb at the QEP, 
these two responses are the most important ones of the $TL$-type. 
The spin-orbit effects are 20\%, 60\% and 75\% of the
response computed with only zeroth-order plus convection terms
in QE conditions at the peak.

Similar trends as those found in the unpolarized $TL$ case
are observed for the rest of the TL-responses
although each one of them presents 
particular features for different kinematics and different values of
$\cal JM$. For obvious reasons it is not possible to show 
all of the figures here; the complete set is presently available from the
authors at \cite{www97}.

We now discuss the TT responses very briefly because they are also in
general very small and hard to measure. 
The unpolarized response is mainly due to the zeroth-order magnetization
term, 
the convection current contributing only a small fraction to 
this response. The PWIA response is completely due
to the convection current and the response computed in PWIA 
is much larger than the effect of the convection upon the DWIA response, 
which implies that the pure convection current is significantly suppressed
by the FSI. Also the response without the SOP is more similar
to a pure reduction of the convection PWIA results than with it.
This suggest that the SOP significantly modifies the cancellation of the
pure magnetization response. And in fact, when a 
calculation was made of the different contributions to the $TT$ responses 
with the SOP set to zero we found that the magnetization
contribution goes away for all of the $TT$ responses, and
not only for the unpolarized one. 
In the other responses of the $TT$-type, we find that the magnetization
contribution is in general dominant; yet in many cases, the convection
contribution is important and several times crucial to 
describe these responses properly. The effect of the convection current 
in general decreases with $q$.

More  interesting is the effect of relativistic corrections upon
the fifth response function.
In general the first-order terms (spin-orbit and convection contributions)
are found to be negligible 
in the QEP region and above it, although there is a noticeable 
effect below the QEP in regions Ia and IIa where 
the fifth response function has two maxima at $p\sim 80$
and 220 MeV/c, and one (positive) minimum in between at $p\sim 130$ MeV/c
(see Fig.~\ref{TL'00}). The effect of the convection current is to 
decrease the response for low $p$
and to increase it for high $p$. The spin-orbit charge increases this
response for low $p$, while  it slightly  decreases it for
high $p$. The effects are noticeable:
for instance, in region IIa the two maxima would be approximately
equal if only magnetization is present. If 
the convection also contributes, then the second peak is bigger than 
the first one by a factor 1.5. If the SOC is also present, then 
the difference between the first and second maxima is about 20\%;
that is the situation shown in Fig.~\ref{TL'00}, where the complete
current is used.
Of course, the FSI effects play an important role here and the
situation is completely changed  if the SOP is zero.

Finally, we have found non-negligible first-order contributions 
to some of the remaining TL$'$ responses, the most important case being the 
$W^{TL'(-)}_{10}$ and
$W^{TL'(-)}_{20}$ responses for a variety of different kinematics.

%=====================
\section{Conclusions}
%=====================

In this work we have investigated polarization observables in coincidence
electron scattering from nuclei when both beam and target are polarized. 
Our primary objective has been to develop the formalism for the 
$\vec{A}(\vec{e},e'N)B$ reaction, introducing a complete 
set of reduced response functions which are independent of the
polarization angles and azimuthal nucleon emission angle, and which contain
the maximum information about nuclear structure and reaction mechanisms of
interest. We have applied the above formalism to the particular case of a 
one-hole nucleus in the shell model using both PWIA and DWIA, and 
we have examined theoretically the 
differences which arise when the nucleon is ejected from different
complete or incomplete shells, and when the daughter nucleus is left in 
states with different angular momenta. With respect to the theoretical
 results for the extreme shell model it is worth pointing out the 
following:

\begin{enumerate}

\item The reduced response functions for nucleon knockout from the same
shell 
but different spin for the daughter nucleus are proportional to the 
responses of a single polarized particle in a shell. The proportionality
 constant  contains a recoupling coefficient of the angular
momenta involved.
When a more sophisticated nuclear model with configuration 
mixing is used, interferences between different shells are
expected \cite{Cab94} and deviations from this proportionality 
could provide information about the nuclear structure involved.

\item Nucleon knockout from a complete inner shell in a polarized 
nucleus can show polarization effects although the involved 
shell has spin zero. The reason is that the final nuclear state 
is left in a state with definite angular momentum, and therefore the 
remaining incomplete shell is partially polarized.
When the sum over final daughter spin is performed, 
all of the polarization responses add to zero. Again, deviations
from this result arise in a more realistic model and, for instance, for a 
deformed nucleus such as $^{21}$Ne, it has been found in Ref.~\cite{Cab94}
using PWIA for the ejected proton that as more states of the 
residual nucleus are involved, the effect of the
target polarization is weakened.

\item The two above results are also valid for PWIA with the
addition that similar conclusions are obtained for the 
polarized momentum distribution of the nucleus: it is
proportional to the polarized momentum distribution of a single 
hole in the shell from which the nucleon is ejected,
with the same proportionality coefficient as found 
between the reduced response functions. Within the context of
this approximation, the study of this reaction
opens the possibility of measuring the complete 
tri-dimensional momentum distribution of the nucleus.

\end{enumerate}

In this work we have focused on the study of the angular reduced reponse 
functions because they are the basic quantities that describe the reaction.
Nevertheless, the large number of these functions (72 for the 
case of $^{39}$K studied here) makes it very difficult to attempt a
separation 
of all of them experimentally, although, of course, specific combinations
of 
the reduced responses could be isolated by performing a combination of 
asymmetry measurements. To understand the relative importance of these 
combinations in (the usual) circumstances where they are not entirely 
isolated it is important to perform an analysis of all of the response 
functions and to study their sensitivities to details of 
the reaction mechanism prior to any measurement. 

That is what we have done in this work for the particular case
of the $^{39}$K nucleus described as a hole in the $d_{3/2}$
shell of $^{40}$Ca. For the ejected proton we have used the DWIA with
an optical potential and compared the results obtained with the PWIA. 
We have included relativity in the kinematics and in the electromagnetic 
current.

We have studied the model dependence of the reduced response functions
by intercomparison between models, specifically addressing the question of 
what is the role of the central and spin-orbit pieces of
the optical potential 
and exploring the effects of
 the first-order pieces of the electromagnetic current
 (convection and spin-orbit charge) in the responses. 
Let us summarize the main conclusions that may be drawn 
from our work.

\begin{enumerate}

\item In general the FSI produce a reduction of the absolute 
strength of the unpolarized responses with respect to the PWIA.
The percentage of this reduction depends on the kinematic
conditions, being typically of $\sim 35\%$
for the $L$ and $T$ responses and for $q=500$ MeV/c at QE conditions,
however being more than a 50\% effect for the $TL$ response. 
The reason of this big reduction is because the FSI generate 
a nonzero contribution from the charge-magnetization,
zeroth-order interference term, which is exactly zero in PWIA, but is
 negative with FSI for $q\ge 500$ MeV/c. We have checked that this effect
 is produced mainly by the spin-orbit potential, since the 
zeroth-order terms go away when we set the SOP equal to zero.
The $TT$ response is a few percent of the others and 
negative in PWIA, but again the FSI produce a drastic
change, even inverting the sign, mainly due to the 
zeroth-order magnetization term in the current,
which is zero in PWIA, although not in DWIA when the SOP is turned on.

\item The fifth response function 
$W^{TL'(+)}_{00}$
which arises for unpolarized nuclei 
and polarized electrons is comparable in size with the 
$W^{TL(+)}_{00}$ response, being around 20--90\%
of the longitudinal unpolarized response under 
different kinematics, and it is 
quite sensitive to the SOP, the most important effects 
being seen below the QEP.

\item In general the FSI effects are large in those responses that are
nonzero in PWIA, 
especially the quadrupole $L$, $T$ and $TL$, and vector and octupole
$T'$ and $TL'$ cases; there is dependence on the particular rank 
$\cal JM$ of each response and in some cases the FSI even produce a 
change of sign. The variety of effects found is too large 
to be summarized in a few lines, but we can cite as most
outstanding cases with dramatic dependence on FSI
the responses
$W^{L(+)}_{21}$,
$W^{T(+)}_{21}$,
$W^{TL(+)}_{20}$,
$W^{TL(+)}_{21}$,
$W^{T'(-)}_{11}$,
$W^{T'(-)}_{3\cal M}$,
$W^{TL'(-)}_{10}$ and 
$W^{TL'(-)}_{3\cal M}$.
Also some cases have been found where the FSI are
negligible, therefore providing an ideal place to 
study other reaction issues, such as those involving the nucleon current 
or details of nuclear structure.
The octupole responses
$W^{T'(-)}_{3\cal M}$ and
$W^{TL'(-)}_{3\cal M}$ are found to be large for ${\cal M}=0,1$, 
especially
for high $q$, and these responses are dominant over the others 
below and above the QEP, probably making their extraction 
experimentally feasible.

\item The polarized responses which are zero in PWIA
are {\it a fortiori} sensitive to the FSI. Additionally, the 
$\widetilde{W}^{L(-)}_{11}$ response is rather dependent on the SOP,
although the 
$\widetilde{W}^{T(-)}_{11}$ response is not.
The effects of the SOP are dramatic in  
$W^{TL(-)}_{1\cal M}$
due to the fact that zeroth-order terms in the current only
contribute
to this response if the SOP is nonzero. Also the 
$W^{TL'(+)}_{20}$
is very sensitive to the SOP.

\item All of the $T$ and $T'$ responses are largely insensitive
to the spin-orbit interaction; therefore they could be used to 
study the central part of the FSI. On the other hand, the $L$,
$TL$ and $TL'$ responses are much more sensitive to the SOP, especially in 
the case of the $TL$ responses.

\item The spin-orbit first-order term in the charge operator
is crucial for properly describing the $TL$ responses
and some of the $TL'$ ones, especially 
at high $q$ where relativistic effects are expected to be more
important. At $q=700$ MeV/c the spin-orbit and convection contribution
are of the same order for both polarized and unpolarized $TL$ responses. 
In other cases the first-order effects are smaller, but noticeable in some
situations, for instance, in the fifth response function.

\item The non-locality correction with a Perey-like factor is in
general smaller than the effects mentioned above and it can be 
safely ignored in a first exploratory study of this kind.

\end{enumerate}

In conclusion, we have found a wide variety of effects
of the FSI and relativistic corrections in the different
reduced response functions of polarized nuclei. 
As the effects are different for different responses, systematic
intercomparisons between selected separated responses
 could provide important
information about different aspects of the reaction mechanism.
The quantity of information which may be obtained
with exclusive polarized electron scattering from polarized
nuclei is more than one order of magnitude
richer than in the unpolarized case and this should allow 
much more stringent constraints 
to be placed on any particular model of the dynamics. 

Experimentally, however, the amount of work involved in measuring
and extracting all of the angular 
reduced response functions is daunting, with 
little hope for a complete extraction of
all of them in the near future, and therefore
it is desirable to extend this analysis to 
the study of spin observables which can be directly measured,
such as cross sections and asymmetries. Our aim here 
has been to provide a first detailed insight into the 
basic quantities describing the reaction, namely the reduced
response functions, as a first step towards understanding
the role of the different reaction ingredients in the problem.
It is our intent to apply the present formalism to study 
various classes of asymmetries and investigate their utility for 
extracting fundamental properties of polarized nuclei such as 
polarized momentum distributions and
spin distributions. Work along these lines is in progress.

%========================================================

\vspace{0.5in}
{\large \bf Acknowledgement}
\vspace{0.1in}

The authors wish to thank Dr. Sabine Jeschonnek for her helpful
comments on the manuscript.

\vspace{0.2in}

%========================================================

%========================================================

\appendix

%========================================================

\renewcommand{\thesection}{Appendix \Alph{section}.}
\renewcommand{\thesubsection}{\Alph{section}.\arabic{subsection}.}

%====================================================

\section{Calculation of the longitudinal response}

%====================================================

Here we present a proof of Eqs.~(\ref{55},\ref{61},\ref{62})
for the longitudinal response. 
We begin from the definition, Eq.~(\ref{22}), and inserting the multipole
expansion in Eq.~(\ref{28}), we have
\begin{equation}
{\cal R}^L=\sum\rho^*\rho = 4\pi\sum_{JJ'}i^{J-J'}[J][J']B^{00}_{J'J},
\label{app188}
\end{equation}
where $B^{00}_{J'J}$ is given by Eq.~(\ref{30}), 
with $\hat{T}'_{J'}=\hat{M}_{J'}$ and $\hat{T}_J=\hat{M}_J$.
Using the general result in Eq.~(\ref{37}) we obtain
\begin{equation}
{\cal R}^L =
        4\pi\sum_{\sigma\sigma'}
        \sum_{\Jb\Jb'L}
        i^{J-J'}i^{l'-l}
        P^+_{l+l'+\cal J'}f^i_{\cal J}
        [Y_{\cal J}Y_{\cal J'}]_{L0}
        \Phi_{\sigma'\sigma}
        \tresj{J}{J'}{L}{0}{0}{0}
        C^*_{\sigma'}C_{\sigma},
\end{equation}
where the Coulomb amplitudes in Eq.~(\ref{40})
have been introduced. Now we use the fact that the 
Coulomb operator $\hat{M}_J$ has natural parity $(-1)^J$. If we denote
by $\Pi_i$ and $\Pi_B$ the parities of the target and daughter
nuclei, respectively, we conclude that the matrix element
$C_{\sigma}$ is nonzero only for transitions satisfying
 $\Pi_i\Pi_B(-1)^{l+J}=1$,
while for $C_{\sigma'}$ we have         
 $\Pi_i\Pi_B(-1)^{l'+J'}=1$. Therefore,
for the product $C_{\sigma'}^*C_{\sigma}$ we have 
the selection rule $l+J+l'+J'=\mbox{even}$.
Taking this into account and using the parity 
function in Eq.~(\ref{51}), we can write
\begin{equation}
i^{J-l-(J'-l')}
C^*_{\sigma'}C_{\sigma}
= \xi^+_{J'-l',J-l}
C^*_{\sigma'}C_{\sigma}
\end{equation}
and we have for the longitudinal response
\begin{equation}
{\cal R}^L =
        4\pi\sum_{\sigma\sigma'}
        \sum_{\Jb\Jb'L}
        P^+_{l+l'+\cal J'}f^i_{\cal J}
        [Y_{\cal J}Y_{\cal J'}]_{L0}
        \Phi_{\sigma'\sigma}
        \tresj{J}{J'}{L}{0}{0}{0}
        \xi^+_{J'-l',J-l}
        C^*_{\sigma'}C_{\sigma}.
        \label{B4}
\end{equation}
Now, using $\xi^+_{mn}=\xi^+_{nm}$, we can write the sum over
$\sigma'\sigma$ in a symmetrized way
\begin{eqnarray}
\lefteqn{\sum_{\sigma\sigma'}
        P^+_{l+l'+\cal J'}
        \Phi_{\sigma'\sigma}
        \tresj{J}{J'}{L}{0}{0}{0}
        \xi^+_{J'-l',J-l}
        C^*_{\sigma'}C_{\sigma}
        =}
        \nonumber\\
&& \frac12\sum_{\sigma\sigma'}
        P^+_{l+l'+\cal J'}
        \tresj{J}{J'}{L}{0}{0}{0}
        \xi^+_{J'-l',J-l}
        \left[  \Phi_{\sigma'\sigma}
                C^*_{\sigma'}C_{\sigma}
                +\Phi_{\sigma\sigma'}
                C^*_{\sigma}C_{\sigma'}
        \right],
        \label{B5}
\end{eqnarray}
because, due to the parity functions, 
$l+l'+{\cal J'}=\rm even$, 
$l+l'+J+J'=\rm even$,
and due to the parity properties of the 3-$j$, we have that 
$J+J'+L=\rm even$.
Using now the symmetry property in Eq.~(\ref{36}),
 we have
$\Phi_{\sigma\sigma'}=(-1)^{\cal J}\Phi_{\sigma'\sigma}$ 
because, due to the parity functions, we have that 
$\Jb'+L=\mbox{even}$. Therefore 
we can write the symmetrized product as
\begin{eqnarray}
\frac12 \left[  \Phi_{\sigma'\sigma}
                C^*_{\sigma'}C_{\sigma}
                +\Phi_{\sigma\sigma'}
                C^*_{\sigma}C_{\sigma'}
        \right]
&=& 
\frac12 \Phi_{\sigma'\sigma}
        \left[  C^*_{\sigma'}C_{\sigma}
                +(-1)^{\cal J}
                C^*_{\sigma}C_{\sigma'}
        \right]
        \nonumber\\
&=& 
        \Phi_{\sigma'\sigma}
        \left[
                P^+_{\cal J}R^L_{\sigma'\sigma}
                +iP^-_{\cal J}I^L_{\sigma'\sigma}
        \right],
        \label{B6}
\end{eqnarray}
where we have used the definition in Eq.~(\ref{43}),
and noticed that when $\Jb=\mbox{even}$, we are taking 
twice the real part of 
$C^*_{\sigma'}C_{\sigma}$,
and when $\cal J =\mbox{odd}$, we have twice the 
imaginary part 
of $C^*_{\sigma'}C_{\sigma}$
times the imaginary unit.

On the other hand, we have the conjugation property
\begin{equation}
[Y_{\cal J}Y_{\cal J'}]^*_{L0}
=(-1)^{\Jb+\Jb'+L}
[Y_{\cal J}Y_{\cal J'}]_{L0}
=(-1)^{\Jb}
[Y_{\cal J}Y_{\cal J'}]_{L0}
\end{equation}
because again $\Jb'+L=\mbox{even}$, and therefore
the coupling 
$[Y_{\cal J}Y_{\cal J'}]_{L0}$ is real for 
$\cal J=\mbox{even}$
and imaginary for 
$\cal J=\mbox{odd}$; and so we can write
\begin{equation} \label{B8}
[Y_{\cal J}Y_{\cal J'}]_{L0}
= P^+_{\cal J}A_{\Jb\Jb'L0}
+i P^-_{\cal J}B_{\Jb\Jb'L0}.
\end{equation}
Finally, inserting Eqs.~(\ref{B5},\ref{B6},\ref{B8}) in Eq.~(\ref{B4})
we obtain:
\begin{eqnarray}
{\cal R}^L &=& 
        4\pi\sum_{\sigma'\sigma}
        \sum_{\Jb\Jb'L}
        P^+_{l+l'+\Jb'}f^i_{\cal J}
        \Phi_{\sigma'\sigma}
        \tresj{J}{J'}{L}{0}{0}{0}
        \xi^+_{J'-l',J-l}
        \nonumber\\
&& \times
        \left[ P^+_{\cal J}
                A_{\Jb\Jb'L0}R^L_{\sigma'\sigma}        
              - P^-_{\cal J}
                B_{\Jb\Jb'L0}I^L_{\sigma'\sigma}        
        \right],
\end{eqnarray}
from which it is straightforward to read off 
Eqs.~(\ref{55},\ref{61},\ref{62}). Although the derivations
of the remaining responses are somewhat more involved,
the basic steps for obtaining them are natural extensions 
of the ones given here.

%===================================================

\section{Reduced responses for a one-hole nucleus}

%===================================================

In this appendix we show how to perform the sums over the angular momenta
of the final states in the reduced responses for 
a one-hole nucleus whose
ground state can be described as an extreme one-hole configuration
$|i^{-1}\rangle$. Since the following steps are analogous for each one 
of the reduced responses $W^{K(\pm)}_{\Jb\Jb'L}$,
we only consider the case of the $W^{L(+)}_{\Jb\Jb'L}$
response, given in general by Eq.~(\ref{61})
\begin{equation}
W^{L(+)}_{\Jb\Jb'L}=
\sum_{\sigma\sigma'}
P^+_{l+l'+\cal J'}
\Phi_{\sigma'\sigma}
\tresj{J}{J'}{L}{0}{0}{0}
\xi^+_{J'-l',J-l}
R^L_{\sigma'\sigma}
\end{equation}
with 
$R^L_{\sigma'\sigma}={\rm Re}\,C^*_{\sigma'}C_{\sigma}$.
First consider the case $h\ne i$, when the struck nucleon is
in a shell other than the polarized one. In this case the final states are
\begin{equation}
|f\rangle = \left[
                a^{\dagger}_j
                [b^{\dagger}_hb^{\dagger}_i]_{J_B}
            \right]_{j_f}
            |C\rangle,
\end{equation}
where $|j\rangle\equiv|E'lj\rangle$ is a single particle in the continuum
with  energy $E'$. Here it is useful to define, in addition
to the index
$\sigma=\{l,j,j_f,J \}$, indices
$\gamma=\{l,j,J\}$, 
$\gamma'=\{l',j',J'\}$,
that do not contain the final angular momenta
$j_f$, $j'_f$ which are going  to be summed. With these 
indices, the many-body Coulomb matrix element can be written as
\begin{equation}
C_{\sigma}=\langle f\| \hat{M}_J\| i^{-1}\rangle
= [J_B][j_f](-1)^{J-j-j_h}
  \seisj{j_i}{j_f}{J}{j}{j_h}{J_B}
  c_{\gamma},
\end{equation}
where we denote with a lower-case letter the
Coulomb matrix element for the single-nucleon transition from the shell
$h$ to the continuum:
\begin{equation}
c_{\gamma}=\langle j \| M_J \| h\rangle.
\end{equation} 
We can then write the real part of the product
$C^*_{\sigma'}C_{\sigma}$ in the following way
\begin{equation}
R^L_{\sigma'\sigma} = 
 [J_B]^2[j_f][j'_f](-1)^{J-j-j_h}(-1)^{J'-j'-j_h}
  \seisj{j_i}{j'_f}{J'}{j'}{j_h}{J_B}
  \seisj{j_i}{j_f}{J}{j}{j_h}{J_B}
  r^L_{\gamma'\gamma},
\end{equation}
where now $r^L_{\gamma'\gamma}$ is the real part
of the interference
between single-particle Coulomb multipoles
\begin{equation}
r^L_{\gamma'\gamma}={\rm Re}\,c^*_{\gamma'}c_{\gamma}.
\end{equation}
Therefore we can write the response as 
\begin{eqnarray}
W^{L(+)}_{\Jb\Jb'L}
&=&     \sum_{j_fj'_f}\sum_{\gamma'\gamma}
         P^+_{l+l'+\Jb'}
         \Phi_{\sigma'\sigma}
         \tresj{J}{J'}{L}{0}{0}{0}
         \xi^+_{J'-l',J-l}
        [J_B]^2[j_f][j'_f]
        \nonumber\\
&& \times
        (-1)^{J+J'+j+j'+1}
          \seisj{j_i}{j'_f}{J'}{j'}{j_h}{J_B}
          \seisj{j_i}{j_f}{J}{j}{j_h}{J_B}
          r^L_{\gamma'\gamma}.
        \label{256}
\end{eqnarray}
Now we are interested in performing the sums
over $j_f,j'_f$. Specifically we are going to 
compute the following coefficient:
\begin{equation}
A\equiv \sum_{j_fj'_f}\Phi_{\sigma'\sigma}[j_f][j'_f]
  \seisj{j_i}{j'_f}{J'}{j'}{j_h}{J_B}
  \seisj{j_i}{j_f}{J}{j}{j_h}{J_B}.
\end{equation}
Inserting the definition in Eq.~(\ref{35}) of $\Phi_{\sigma'\sigma}$,
\begin{eqnarray}
A &=& \sum_{j_fj'_f}
        [J][J'][j][j'][j_f][j'_f][\Jb'][L]
        (-1)^{J'+J_B+j_f+1/2+\Jb+\Jb'} 
        \nonumber\\
&&\times
        \tresj{j'}{j}{\cal J'}{\textstyle\frac12}{-\textstyle\frac12}{0}
        \seisj{j'}{j}{\cal J'}{j_f}{j'_f}{J_B}
        \nuevej{J}{J'}{L}{J_i}{J_i}{\cal J}{j_f}{j'_f}{\cal J'}
        \nonumber\\
&&\times
        [j_f][j'_f]
  \seisj{j_i}{j'_f}{J'}{j'}{j_h}{J_B}
  \seisj{j_i}{j_f}{J}{j}{j_h}{J_B}
\end{eqnarray}
The  sum over $j_f$, $j'_f$
 can be written as the product of a 6-$j$ and a 9-$j$
symbol using the relation 
\begin{eqnarray}
\lefteqn{\nuevej{a}{b}{c}{a'}{b'}{c'}{d}{e}{f}
        \seisj{d}{e}{f}{h}{i}{j}=}
        \nonumber\\
&=&     \sum_{\lambda\mu}
        (-1)^{S+\lambda}
        [\lambda]^2[\mu]^2
        \seisj{c}{c'}{f}{h}{i}{\lambda}
        \seisj{a'}{b'}{c'}{h}{\lambda}{\mu}
        \seisj{b}{b'}{e}{h}{j}{\mu}
        \nuevej{a}{b}{c}{a'}{\mu}{\lambda}{d}{j}{i},
        \nonumber\\
\end{eqnarray}
with $S=b+c-a'-c'-d-e+2h+j$.
Note that Rotenberg's version \cite{Rot59} of this
equation (Eq.~(3.23)) has an error in the third 6-$j$
of the righthand side: instead of $h$, in that reference 
appear an $i$. In order to obtain the correct 
expression, in this work we have proven the above
relation using the graphical method of Ref.~\cite{Dan90};
for brevity we do not reproduce the procedures used here.

Using the above relation and after some rearrangement 
and simplication of phases we obtain for the  sum $A$ the result
\begin{eqnarray}
A &=& [J][J'][j][j'][\Jb'][L]
        (-1)^{J'+J_B+\cal J+J'}
        \tresj{j'}{j}{\cal J'}{\frac12}{-\frac12}{0}
        \nonumber\\
&& \times
         (-1)^{J+J'+\cal J+J'}
        (-1)^{j+j_h+j_i-1/2}
        \nuevej{J}{J'}{L}{j_h}{j_h}{\cal J}{j}{j'}{\cal J'}
        \seisj{j_i}{j_i}{\cal J}{j_h}{j_h}{J_B}.
\end{eqnarray}
The 9-$j$ coefficient has the same structure as 
a polarized particle with spin $j_h$ would have. In order to relate
the result with the response of a polarized particle
in the shell $h$ and spin-zero daughter nucleus, $J_B=0$,
we first notice that 
the coefficient $\Phi_{\gamma'\gamma}$
corresponding to Eq.~(\ref{35}) 
for that process can be written
\begin{eqnarray}
\Phi_{\gamma'\gamma}(\Jb\Jb'L,h\rightarrow 0) &=&
        [J][J'][j][j'][\Jb'][L][j][j']
        (-1)^{J'+j+1/2+\cal J+J'}
        \nonumber\\
&&\kern -2cm \times
        \tresj{j'}{j}{\cal J'}{\frac12}{-\frac12}{0}
        \seisj{j'}{j}{\cal J'}{j}{j'}{0}
        \nuevej{J}{J'}{L}{j_h}{j_h}{\cal J}{j}{j'}{\cal J'}.
\end{eqnarray}
Inserting the simple result $(-1)^{j+j'+\cal J'}/[j][j']$
for the 6-$j$ symbol with one argument equal to zero, 
the coefficient $\Phi_{\gamma'\gamma}$ becomes
\begin{eqnarray}
\Phi_{\gamma'\gamma}(\Jb\Jb'L,h\rightarrow 0) &=&
        [J][J'][j][j'][\Jb'][L]
        (-1)^{J'+j+1/2+\cal J+J'}
        \nonumber\\
&& \times
        \tresj{j'}{j}{\cal J'}{\frac12}{-\frac12}{0}
        (-1)^{\Jb'+j+j'}
        \nuevej{J}{J'}{L}{j_h}{j_h}{\cal J}{j}{j'}{\cal J'}.
\end{eqnarray}
Then we obtain for the sum $A$ the value
\begin{equation}
A= (-1)^{J+J'+\Jb+j+j'}(-1)^{j_h+j_i+J_B+1}
        \Phi_{\gamma'\gamma}(h\rightarrow0)
        \seisj{j_i}{j_i}{\cal J}{j_h}{j_h}{J_B}.
\end{equation}
Finally, inserting this value into Eq.~(\ref{256}) we
have for the longitudinal response
\begin{eqnarray}
W^{L(+)}_{\Jb\Jb'L} &=&
        (-1)^{j_h+j_i+J_B+\cal J}
        [J_B]^2
        \seisj{j_i}{j_i}{\cal J}{j_h}{j_h}{J_B}
        \nonumber\\
&& \times
        \sum_{\gamma'\gamma}
        P^+_{l+l'+\cal J'}
        \Phi_{\gamma'\gamma}(h\rightarrow 0)
        \tresj{J}{J'}{L}{0}{0}{0}
        \xi^+_{J'-l',J-l}
        r^L_{\gamma'\gamma}
\end{eqnarray}
from which we directly obtain Eq.~(\ref{77b}).

For the other case, $h=i$, the final hadronic states are
of the type
\begin{equation}
|f\rangle = \frac{1}{\sqrt{2}}\left[
                a^{\dagger}_j
                [b^{\dagger}_ib^{\dagger}_i]_{J_B}
            \right]_{j_f}
            |C\rangle,
\end{equation}
and the transition  matrix element is now
\begin{equation}
C_{\sigma}=\langle f\| \hat{M}_J\| i^{-1}\rangle
=   \sqrt{2}
[J_B][j_f](-1)^{J-j-j_i}
  \seisj{j_i}{j_f}{J}{j}{j_i}{J_B}
c_{\gamma}.
\end{equation}
The rest of the derivation is completely analogous, with 
the exception of the factor $\sqrt{2}$, that becomes 
a factor of two in the response. Therefore, making $h=i$
in the above equations and multiplying by two,
it is straightforward to obtain Eq.~(\ref{84}).

%========================================================

\begin{table}[p]
\begin{center}
\begin{tabular}{lrrrrrrrrrrr} \hline\\ 
$\omega$[MeV]           & 
$\cal J$                & 
$\cal M$                &
$W^L$    & $W^T$        & 
$W^{TL}$                & 
$\widetilde{W}^{TL}$    &
$W^{TT}$                & 
$\widetilde{W}^{TT}$    &
$W^{T'}$                &
$W^{TL'}$               & 
$\widetilde{W}^{TL'}$   \\ \hline\hline
40
& 0& 0& 100.0&  38.9&  53.3&     -&  10.8&     -&     -&  23.8&     -\\
& 1& 0&     -&     -&  30.5&     -&   7.4&     -&  42.0&  16.6&     -\\
& 1& 1&  29.6&  15.3&  17.9&  34.1&   8.3&  10.5&  11.3& 116.2&  78.7\\
& 2& 0& 129.1&  36.0&  29.7&     -&   9.4&     -&     -&  61.6&     -\\
& 2& 1&  24.0&   4.0&  24.1&   7.7&   5.8&   9.6&   3.4&  48.7&  74.3\\
& 2& 2&  13.7&   7.6&   8.6&  10.1&   3.0&   4.2&   8.1&   4.9&  11.7\\
& 3& 0&     -&     -&  15.4&     -&   8.8&     -&  28.1&  52.7&     -\\
& 3& 1&  15.4&   2.1&  17.1&  13.4&   2.7&   1.2&  10.0&  38.1&  39.2\\
& 3& 2&   9.8&   3.1&   4.7&   5.7&   3.6&   2.2&   2.1&   7.0&   5.7\\
& 3& 3&   2.4&   0.3&   2.3&   1.7&   1.0&   1.1&   1.2&   5.8&   5.4\\
\hline
56
& 0& 0& 100.0&  39.8&  53.6&     -&   5.2&     -&     -&  58.2&     -\\
& 1& 0&     -&     -&  24.3&     -&   3.6&     -&  41.5&  36.6&     -\\
& 1& 1&  55.6&  22.2&  12.2&  31.7&   9.1&   8.1&  10.4&  79.3&  70.7\\
& 2& 0&  89.1&  23.5&  36.5&     -&   9.2&     -&     -&  14.1&     -\\
& 2& 1&  19.7&   6.9&   6.8&   3.3&   4.7&   5.7&   7.2&  24.7&  27.8\\
& 2& 2&  18.9&   8.1&  10.8&   3.9&   1.3&   1.3&   3.9&  14.7&  21.4\\
& 3& 0&     -&     -&  17.0&     -&   7.5&     -&  20.5&  69.5&     -\\
& 3& 1&   7.3&   2.9&  12.4&   9.9&   3.6&   5.0&  10.0&  34.1&  21.5\\
& 3& 2&   7.1&   1.8&   5.0&   3.3&   1.4&   1.8&   4.0&   8.9&   5.0\\
& 3& 3&   0.9&   0.3&   2.2&   1.9&   1.0&   1.1&   0.9&   6.4&   6.3\\
\hline
80
& 0& 0& 100.0&  36.8&  45.1&     -&   3.2&     -&     -&  38.4&     -\\
& 1& 0&     -&     -&  12.4&     -&   3.2&     -&  38.6&  86.0&     -\\
& 1& 1&  38.9&  14.6&  13.3&  21.0&   5.2&   3.5&  26.2&  97.0&  93.2\\
& 2& 0&  99.6&  31.9&  28.3&     -&   2.3&     -&     -&  20.2&     -\\
& 2& 1&  49.5&  17.0&  19.7&   0.9&   1.0&   1.0&   5.3&  22.6&  18.5\\
& 2& 2&  17.1&   6.1&   9.6&   0.9&   1.1&   0.5&   2.0&  11.6&  11.6\\
& 3& 0&     -&     -&  40.3&     -&  15.9&     -&  55.5& 147.9&     -\\
& 3& 1&   4.4&   0.8&  23.5&  14.6&   5.3&   7.0&  23.6&  87.1&  41.5\\
& 3& 2&   0.3&   0.3&   5.0&   3.0&   1.5&   3.0&   5.2&  16.6&  12.6\\
& 3& 3&   0.9&   0.3&   1.5&   1.5&   0.5&   0.8&   1.2&   3.5&   4.6\\
\hline
\end{tabular}
\end{center}
\caption{\em Total variation (maximum minus minimum) of the 
reduced response functions as a percentage of the 
variation of the unpolarized longitudinal. The responses
are computed in DWIA for $q=300$ MeV/c.}  
\end{table}

\begin{table}[p]
\begin{center}
\begin{tabular}{lrrrrrrrrrrr} \hline\\
$\omega$[MeV]           & 
$\cal J$                & 
$\cal M$                &
$W^L$    & $W^T$        & 
$W^{TL}$                & 
$\widetilde{W}^{TL}$    &
$W^{TT}$                & 
$\widetilde{W}^{TT}$    &
$W^{T'}$                &
$W^{TL'}$               & 
$\widetilde{W}^{TL'}$   \\ \hline\hline
110.0
& 0& 0& 100.0& 100.1&  43.9&     -&   9.8&     -&     -&  20.8&     -\\
& 1& 0&     -&     -&  16.5&     -&  16.1&     -&  74.1&  89.3&     -\\
& 1& 1&  16.0&  25.7&  14.1&  14.1&  16.2&   5.5&  46.6&  70.0& 165.6\\
& 2& 0&  36.9&  44.1&  19.0&     -&   4.8&     -&     -&  51.8&     -\\
& 2& 1&  25.0&  30.8&  21.0&   5.6&   3.0&   4.2&   6.4&  22.3&  28.3\\
& 2& 2&  22.8&  21.2&   9.7&   2.0&   2.9&   5.2&   7.0&   5.5&   7.3\\
& 3& 0&     -&     -&  16.0&     -&  12.3&     -&  25.6& 135.4&     -\\
& 3& 1&   5.5&   5.5&  13.2&   7.6&   3.1&   4.4&  37.7&  35.8&  17.2\\
& 3& 2&   0.7&   2.7&   1.6&   3.7&   2.5&   2.4&   3.9&  23.4&  12.7\\
& 3& 3&   1.5&   0.3&   1.3&   1.5&   0.8&   1.1&   3.7&   7.9&  
9.6\\\hline
133.5
& 0& 0& 100.0& 103.1&  48.7&     -&   9.1&     -&     -&  37.5&     -\\
& 1& 0&     -&     -&  26.1&     -&   8.5&     -&  96.4&  11.6&     -\\
& 1& 1&  23.9&  32.0&   4.9&  17.9&   9.3&   8.0&  18.2&  97.2& 158.2\\
& 2& 0&  50.4&  57.5&  18.8&     -&   9.1&     -&     -&  19.9&     -\\
& 2& 1&   8.6&   8.6&  13.3&   3.9&   3.4&   5.5&   7.4&  19.6&  17.7\\
& 2& 2&  23.7&  24.4&  12.2&   1.4&   2.6&   3.5&   5.3&  13.4&  14.3\\
& 3& 0&     -&     -&  14.1&     -&   5.5&     -&  61.8&  42.9&     -\\
& 3& 1&   1.8&   1.6&   4.7&   2.6&   1.9&   3.5&  17.6&  67.6&  22.9\\
& 3& 2&   2.2&   2.1&   2.6&   3.2&   1.3&   0.9&  10.4&   6.1&   3.7\\
& 3& 3&   1.0&   0.7&   0.8&   0.6&   1.1&   1.0&   1.8&  11.3& 
11.1\\\hline
160.0
& 0& 0& 100.0& 104.0&  50.5&     -&   4.8&     -&     -&  42.9&     -\\
& 1& 0&     -&     -&  22.5&     -&   3.2&     -& 100.3&  95.4&     -\\
& 1& 1&  29.2&  33.8&  12.4&  19.5&   8.0&   5.4&  53.1& 101.7& 160.1\\
& 2& 0&  53.7&  65.9&  23.6&     -&  11.4&     -&     -&  15.4&     -\\
& 2& 1&  34.5&  38.4&   4.6&   5.7&   6.8&   5.7&   9.9&   7.0&  13.4\\
& 2& 2&  21.5&  22.2&  12.3&   2.7&   2.3&   3.6&   2.2&  15.2&  14.2\\
& 3& 0&     -&     -&  17.1&     -&   9.8&     -&  78.5& 182.7&     -\\
& 3& 1&   2.3&   2.0&  10.4&   7.2&   3.5&   5.3&  49.8&  81.1&  33.3\\
& 3& 2&   1.2&   0.6&   3.0&   2.1&   1.3&   1.7&  13.3&  26.7&  16.3\\
& 3& 3&   0.5&   0.9&   1.2&   1.3&   0.8&   0.9&   3.3&   8.9& 
10.2\\\hline
\end{tabular}
\end{center}
\caption{\em Total variation (maximum minus minimum) of the 
reduced response functions as a percentage of the 
variation of the unpolarized longitudinal. The responses
are computed in DWIA for $q=500$ MeV/c.}  
\end{table}

\begin{table}[p]
\begin{center}
\begin{tabular}{lrrrrrrrrrrr} \hline\\
$\omega$[MeV]           & 
$\cal J$                & 
$\cal M$                &
$W^L$    & $W^T$        & 
$W^{TL}$                & 
$\widetilde{W}^{TL}$    &
$W^{TT}$                & 
$\widetilde{W}^{TT}$    &
$W^{T'}$                &
$W^{TL'}$               & 
$\widetilde{W}^{TL'}$   \\ \hline\hline
180
& 0& 0& 100.0& 162.6&  30.4&     -&   3.6&     -&     -&  22.9&     -\\
& 1& 0&     -&     -&   1.4&     -&   7.3&     -&  94.0& 181.7&     -\\
& 1& 1&   5.9&  31.0&  12.0&  20.1&  12.0&   6.2& 105.2& 177.9& 230.6\\
& 2& 0&  80.9& 112.5&  10.7&     -&  10.1&     -&     -&  38.0&     -\\
& 2& 1&  49.8&  80.0&   9.5&  13.0&   7.3&   6.4&   8.5&  23.6&  12.8\\
& 2& 2&  14.5&  21.2&   8.0&   3.8&   3.7&   3.6&   3.3&   2.3&   1.2\\
& 3& 0&     -&     -&  37.0&     -&  19.3&     -& 246.0& 271.1&     -\\
& 3& 1&   1.2&   6.7&  23.3&  18.3&   8.5&  11.0&  76.4& 187.3&  76.1\\
& 3& 2&   0.4&   1.5&   3.7&   3.4&   3.4&   5.1&  18.9&  29.4&  27.1\\
& 3& 3&   0.2&   0.1&   0.6&   0.7&   0.7&   0.8&   4.9&   4.1&  
7.5\\\hline
241
& 0& 0& 100.0& 177.0&  45.7&     -&   6.2&     -&     -&  80.4&     -\\
& 1& 0&     -&     -&  14.2&     -&  12.8&     -& 164.8&  42.0&     -\\
& 1& 1&  43.7&  80.1&  10.4&  32.5&   5.8&   3.3&  29.7& 117.4& 212.9\\
& 2& 0&  53.2&  99.5&  22.8&     -&   3.9&     -&     -&   5.7&     -\\
& 2& 1&   5.0&  15.6&   7.8&   0.5&   4.4&   6.2&  20.6&  11.9&   6.5\\
& 2& 2&  23.5&  41.0&  12.0&   1.7&   2.0&   2.0&   5.5&  28.9&  19.3\\
& 3& 0&     -&     -&   3.4&     -&  10.0&     -&  89.9&  43.2&     -\\
& 3& 1&   2.1&   5.2&  11.4&   2.5&   0.5&   0.4&  26.0&  72.3&  28.4\\
& 3& 2&   0.9&   2.0&   0.5&   1.5&   1.8&   1.9&  15.8&   7.8&   4.5\\
& 3& 3&   0.7&   2.2&   0.7&   0.9&   0.5&   0.5&   2.8&  14.9& 
15.0\\\hline
300
& 0& 0& 100.0& 188.4&  40.1&     -&   5.2&     -&     -&  91.0&     -\\
& 1& 0&     -&     -&  13.5&     -&   1.9&     -& 174.5& 170.8&     -\\
& 1& 1&  52.4&  86.6&  10.4&  38.5&   6.3&   3.0& 127.4& 206.8& 223.9\\
& 2& 0&  80.7& 177.1&   9.4&     -&  15.9&     -&     -&  59.4&     -\\
& 2& 1&  47.3&  92.4&   7.7&  10.8&  10.1&   8.3&  19.8&  33.4&  40.8\\
& 2& 2&  16.4&  31.2&  11.1&   4.1&   5.3&   5.9&  11.2&  26.3&  16.2\\
& 3& 0&     -&     -&  16.3&     -&   4.8&     -& 294.0& 335.3&     -\\
& 3& 1&   6.0&   9.7&   8.9&   4.6&   1.3&   3.0& 118.5& 195.1&  93.4\\
& 3& 2&   2.4&   5.1&   1.8&   0.9&   0.8&   0.6&  25.1&  38.2&  30.0\\
& 3& 3&   0.6&   1.7&   0.7&   1.1&   0.5&   0.5&   6.1&   8.1& 
10.8\\\hline
\end{tabular}
\end{center}
\caption{\em Total variation (maximum minus minimum) of the 
reduced response functions as a percentage of the 
variation of the unpolarized longitudinal. The responses
are computed in DWIA for $q=700$ MeV/c.}  
\end{table}

\clearpage

%   *********************
%   *                   *
%   *  Figure caption   *
%   *                   *
%   *********************

\newpage

\begin{figure}
\caption{
\label{L00}
Response function 
$W^{L(+)}_{00}$
shown as a function of the missing momentum $p$ for the nine
kinematical conditions specified in Table~1.
Solid lines: computed 
with the complete optical potential; dashed lines:
computed with the spin-orbit potential equal to zero;
dot-dashed lines: computed with the total potential and with 
Perey parameter $\beta=0.85$ in the wave function;
dotted curves: PWIA.
}
\end{figure}

\begin{figure}

\caption{
\label{L11tilde} The same as Fig.~\ref{L00}, but for the 
vector longitudinal response 
$\widetilde{W}_{11}^{L(-)}$
}
\end{figure}

\begin{figure}
\caption{
\label{L20} The same as Fig.~\ref{L00}, but for the 
quadrupole longitudinal response 
${W}_{20}^{L(+)}$
}
\end{figure}

\begin{figure}
\caption{
\label{T00} The same as Fig.~\ref{L00}, but for the 
unpolarized transverse response 
${W}_{00}^{T(+)}$
}
\end{figure}

\begin{figure}
\caption{
\label{TL00} The same as Fig.~\ref{L00}, but for the 
unpolarized transverse-longitudinal response 
${W}_{00}^{TL(+)}$
}
\end{figure}

\begin{figure}
\caption{
\label{TL10} The same as Fig.~\ref{L00}, but for the 
vector $TL$ response 
${W}_{10}^{TL(-)}$
}
\end{figure}

\begin{figure}
\caption{
\label{TL20} The same as Fig.~\ref{L00}, but for the 
quadrupole $TL$ response 
${W}_{20}^{TL(+)}$
}
\end{figure}

\clearpage

\begin{figure}
\caption{
\label{T'10} The same as Fig.~\ref{L00}, but for the 
vector T' response 
${W}_{10}^{T'(-)}$
}
\end{figure}

\begin{figure}
\caption{
\label{T'11} The same as Fig.~\ref{L00}, but for the 
vector T' response 
${W}_{11}^{T'(-)}$
}
\end{figure}

\begin{figure}
\caption{
\label{T'30} The same as Fig.~\ref{L00}, but for the 
octupole T' response 
${W}_{30}^{T'(-)}$
}
\end{figure}

\begin{figure}
\caption{
\label{T'31} The same as Fig.~\ref{L00}, but for the 
octupole T' response 
${W}_{31}^{T'(-)}$
}
\end{figure}

\begin{figure}
\caption{
\label{TL'00} The same as Fig.~\ref{L00}, but for the 
fifth response function 
${W}_{00}^{TL'(+)}$
}
\end{figure}

\begin{figure}
\caption{
\label{TL'10} The same as Fig.~\ref{L00}, but for the 
vector $TL'$  response 
${W}_{10}^{TL'(-)}$
}
\end{figure}

\begin{figure}
\caption{
\label{TL'20} The same as Fig.~\ref{L00}, but for the 
quadrupole $TL'$ response 
${W}_{20}^{TL'(+)}$
}
\end{figure}

\begin{figure}
\caption{
\label{TL'30} The same as Fig.~\ref{L00}, but for the 
octupole $TL'$ response 
${W}_{30}^{TL'(-)}$
}
\end{figure}

\begin{figure}
\caption{
\label{soTL00}
Response function 
$W^{TL(+)}_{00}$ computed in DWIA with the complete optical potential,
shown as a function of the missing momentum $p$, for the nine
kinematical conditions specified in Table~1.
Solid lines: total response  including all the pieces of the
electromagnetic current;
dashed lines: only including charge plus magnetization 
(zeroth-order) terms; dot-dashed lines: including, in addition,
 the convection current.
}
\end{figure}

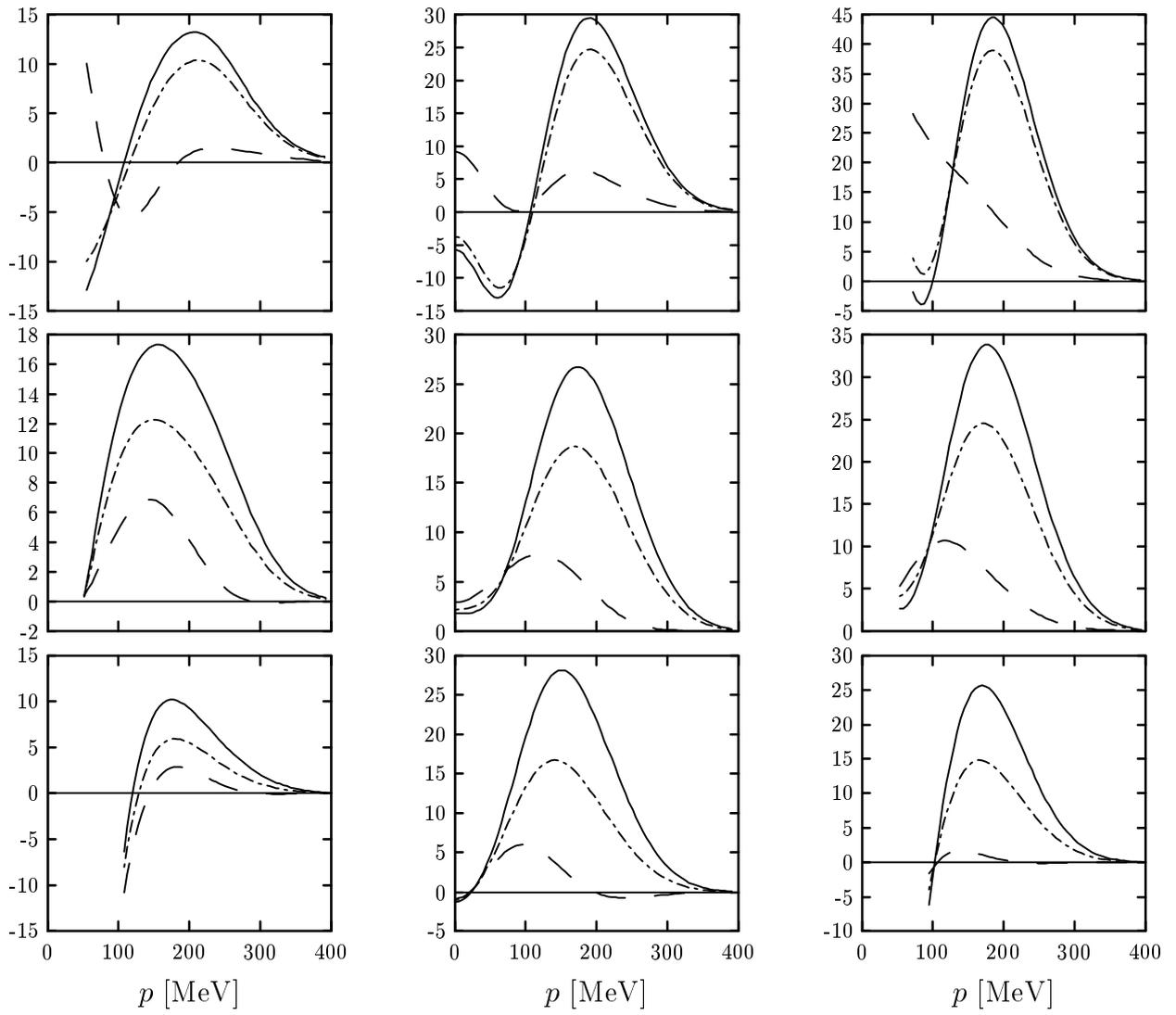
\begin{figure}
\caption{
\label{soTL11tilde} The same as Fig.~\ref{soTL00}, but for the 
vector $TL$ response 
$\widetilde{W}_{11}^{TL(-)}$.
}
\end{figure}

\end{document}